\documentclass[twocolumn]{openjournal}
\usepackage{graphicx} 
\usepackage{multirow}

\usepackage[colorlinks,linkcolor=blue,citecolor=blue,urlcolor=blue ]{hyperref}
\usepackage[utf8]{inputenc}
\usepackage{float}
\usepackage{xcolor}
\usepackage{ulem}
\usepackage[T1]{fontenc}

\def\sigh2{$\Sigma_{\rm H_2}$}

\def\kms{km~s$^{-1}$}

\begin{document}
\title{Galaxy evolution in the post-merger regime.  III - The triggering of active galactic nuclei peaks immediately after coalescence\vspace{-1.5cm}}

\author{Sara L. Ellison$^1$}
\author{Leonardo Ferreira$^1$}
\author{Robert Bickley$^1$}
\author{Tess Grindlay$^1$}
\author{Samir Salim$^2$}
\author{~~~~~~~~~~~~~~~~Shoshannah Byrne-Mamahit$^1$}
\author{Shobita Satyapal$^3$}
\author{David R. Patton$^4$}
\author{Jillian M. Scudder$^5$}
\thanks{$^*$E-mail: sarae@uvic.ca}
\affiliation{$^1$ Department of Physics \& Astronomy, University of Victoria, Finnerty Road, Victoria, BC V8P 1A1, Canada\\
  $^2$ Department of Astronomy, Indiana University, Bloomington, Indiana 47405, USA\\
  $^3$ George Mason University, Department of Physics and Astronomy, MS3F3, 4400 University Drive, Fairfax, VA 22030, USA\\
  $^4$ Department of Physics and Astronomy, Trent University, 1600 West Bank Drive, Peterborough, ON K9L 0G2, Canada\\
  $^5$ Department of Physics and Astronomy, Oberlin College, Oberlin, Ohio, OH 44074, USA
}

\begin{abstract}

Galaxy mergers have been shown to trigger active galactic nuclei (AGN) in the nearby universe, but the timescale over which this process happens remains unconstrained.  The Multi-Model Merger Identifier (\textsc{mummi}) machine vision pipeline has been demonstrated to provide reliable predictions of time post-merger ($T_{PM}$) for galaxies selected from the Ultraviolet Near Infrared and Optical Northern Survey (UNIONS) up to $T_{PM}$=1.76 Gyr after coalescence.  By combining the post-mergers identified in UNIONS with pre-coalescence galaxy pairs, we can, for the first time, study the triggering of AGN throughout the merger sequence.  AGN are identified using a range of complementary metrics: mid-IR colours, narrow emission lines and broad emission lines, which can be combined to provide insight into the demographics of dust and luminosity of the AGN population.  Our main results are:  1) Regardless of the metric used, we find that the peak AGN excess (compared with a matched control sample) occurs immediately after coalescence, at $0 < T_{PM} < 0.16$ Gyr.  2)  The excess of AGN is observed until long after coalescence; both the mid-IR selected AGN and broad line AGN are more common than in the control sample even in the longest time bin of our sample ($0.96 < T_{PM}$ < 1.76 Gyr).  3)  The AGN excess is larger for more luminous and bolometrically dominant AGN, and we find that AGN in post-mergers are generally more luminous than secularly triggered events.  4)  A deficit of broad line AGN in the pre-merger phase, that evolves into an excess in post-mergers is consistent with evolution of the covering fraction of nuclear obscuring material.  Before coalescence, tidally triggered inflows increase the covering fraction of nuclear dust; in the post-merger regime feedback from the AGN clears (at least some of) this material.  5)  The statistical peak in the triggering of starbursts occurs contemporaneously with AGN, within 0.16 Gyr of coalescence.
  
\end{abstract}
\maketitle

\section{Introduction} \label{intro_sec}

For many years, the link between galaxy mergers and the triggering of nuclear accretion was considered highly contentious.  Indeed, it had become an almost customary component of any introduction on the subject to point out the slew of papers that both found support for (e.g. Alonso et al. 2007; Woods \& Geller 2007; Koss et al. 2010; Ellison et al. 2011; Liu et al. 2012; Sabater et al. 2013; Khabiboulline et al. 2014; Satyapal et al. 2014) and against (e.g. Ellison et al. 2008; Li et al. 2008; Gabor et al. 2009; Darg et al. 2010; Cisternas et al. 2011; Scott \& Kaviraj 2014) a link between galaxy mergers and a statistical excess of active galactic nuclei (AGN).  Since these pioneering works of over a decade ago, we have learned many lessons, including the importance of large statistical datasets, rigourously matched control samples and multi-wavelength AGN selection.  Thus, although the situation at $z>0.5$ is still debated, with some studies supporting a connection between mergers and AGN (e.g. Silverman et al. 2011; Dougherty et al. 2024; La Marca et al. 2024) and others not (e.g. Kocevski et al. 2012; Shah et al. 2020; Silva et al. 2021; Lambrides et al. 2021), modern studies of low redshift galaxies ($z<0.5$) now generally agree that mergers are capable of triggering AGN (e.g. Weston et al. 2017; Goulding et al. 2018; Ellison et al. 2019; Gao et al. 2020; Duplancic et al. 2021; Bickley et al. 2023, 2024a; Li et al. 2023; La Marca et al. 2024).

\smallskip

Despite the broad consensus view that low $z$ mergers can trigger AGN, many details of the merger-AGN connection remain unclear.  For example, whether or not merger triggered AGN are more luminous (Koss et al. 2012; Satyapal et al. 2014; Marian et al. 2020; Bickley et al. 2023; Villforth et al. 2017; Villforth 2023; La Marca et al. 2024) or more obscured (Kocevski et al. 2015; Ricci et al. 2017; Satyapal et al. 2017; Koss et al. 2018; Li et al. 2020; Barrows et al. 2023) than secularly triggered AGN, and whether or not feedback is capable of excavating nuclear gas and dust in a so-called blowout phase (e.g. Hou et al. 2020, 2023; Ricci et al. 2021; Bickley et al. 2024a).  Definitive answers to these questions remain elusive, with the added complexity that factors such as galaxy mass, star formation rate or the mass ratio of the merger may influence the outcome (e.g. Stemo et al. 2021; Barrows et al. 2023).  What seems clear, however, is that the details of the sample selection, both in terms of galaxy (e.g. stellar mass, redshift, orbital geometry and merger mass ratio) and AGN properties (e.g. luminosity and obscuration) all affect the extent to which mergers and AGN are connected, as also found in simulations (e.g. Capelo et al. 2015; Blecha et al. 2018; Steinborn et al. 2018; McAlpine et al. 2020).  

\smallskip

In addition to the many galaxy-centric factors (such as gas fraction and morphology) that impact the triggering of AGN during galaxy interactions, another important variable that is expected to influence experimental outcomes is time along the merger sequence.   On the one hand, the time evolution of galaxy mergers is a natural axis to investigate with simulations and several works have reported a consensus view that elevated accretion can begin in the encounter phase, but peaks around the time of coalescence (Capelo et al. 2015; McAlpine et al. 2020; Byrne-Mamahit et al. 2023).  Most recently, by tracking black hole accretion rates throughout the merger sequence in a large sample drawn from the Illustris-TNG simulation, Byrne-Mamahit et al. (2024) found that the statistical elevation of merger triggered nuclear accretion is long lived, persisting for $\sim$ 2 Gyr post-coalescence.

\smallskip

On the other hand, studying merger properties as a function of time is extremely challenging for observational works.  The popular proxy for time in the pre-coalescence regime is the projected separation ($r_p$) between pairs of galaxies, which is statistically well motivated by simulations (e.g. Patton et al. 2024).  Many observational works have found the greatest enhancement in AGN activity at small pair separations (e.g. Ellison et al. 2011; Khabiboulline et al 2014; Hou et al. 2020; Stemo et al. 2021), motivating studies to push into the post-coalescence regime.  Indeed, studies of post-mergers have found that the highest rates of AGN activity occur after coalescence (Ellison et al. 2013; Satyapal et al. 2014; Bickley et al., 2023, 2024a), in agreement with simulations.  However, what remains completely unconstrained in the observations is how the frequency of AGN evolves with time after coalescence, both in terms of the epoch at which it peaks, and for how long statistical enhancements persist.  The application of artificial intelligence to not only identify mergers (Pearson et al. 2019, 2022; Ferreira et al. 2020;  Bickley et al. 2021; Walmsley et al. 2022; Omori et el. 2023) but also predict the time since coalescence of a given post-merger remnant, is just now becoming feasible (Koppula et al. 2021; Pearson et al. 2024).

\smallskip

In order to study the effects of galaxy interactions as a function of time post-merger ($T_{PM}$) we have recently developed the MUlti Model Merger Identifier (\textsc{mummi}), a machine vision based framework that can both find mergers and predict $T_{PM}$ (Ferreira et al. 2024; Ferreira et al. in prep).  In our first two papers in the science series, we have used \textsc{mummi} to investigate the post-coalescence time evolution of star formation triggering (Ferreira et al. 2025) and quenching (Ellison et al. 2024).  In the work presented here we turn our attention to AGN, in order to assess both the epoch of peak nuclear accretion as well as the duration over which the merger influences the heightened activity.

\section{Data}\label{data_sec}

The goal of the current work is to track the timescale of AGN triggering during galaxy interactions, as well as to assess any luminosity dependence and the evolution of dust obscuration.  In order to achieve these objectives, we must assemble samples of both pre-coalescence and post-merger galaxies, as well as control samples for each.  AGN must then be identified in the various samples so that we can assess their relative frequency in mergers and control galaxies. In this section we describe the compilation of these various datasets and the methods by which AGN are identified.

\subsection{Galaxy metadata}\label{pair_sec}

Redshifts and emission line fluxes are taken from the SDSS DR7, as measured in the MPA-JHU catalogs\footnote{https://home.strw.leidenuniv.nl/$\sim$jarle/SDSS/}.  We apply dust corrections to all catalog emission lines by adopting a Small Magellanic Cloud (SMC) extinction curve (Pei 1992) and assuming an intrinsic Balmer decrement of 2.85.  In contrast to most of our previous work that uses total stellar masses from the MPA-JHU catalog (Kauffmann et al. 2003a), here we re-compute our own stellar masses using SDSS $griz$ photometry.  The reason for this change is that the MPA-JHU catalog does not include stellar masses for galaxies with broad emission lines, which represent an important AGN identification technique.  Our method follows closely the one laid out in Salim et al. (2016), but without the use of the bluest bands, where we expect AGN contamination can be significant.  Of the galaxies in common with the MPA/JHU catalog, almost 90 per cent have stellar masses that agree to within 0.1 dex with our new derivations, with a median difference of only 0.03 dex.  More details on the photometrically derived stellar masses, comparisons with the MPA-JHU stellar masses and robustness of the limited $griz$ filter set are all presented and discussed in Bickley et al. (2024a).

\subsection{The galaxy pair sample}\label{pair_sec}

Our pairs sample is constructed following the methods laid out in Patton et al. (2016), except that we use our new photometrically derived stellar masses that include galaxies with a possible broad line component.  The pairs sample is compiled by identifying the closest companion of each galaxy in the SDSS DR7 whose stellar mass is within a factor of 10, whose velocity separation $\Delta v < 300$ \kms\ and whose projected separation is $r_p < 100$ kpc.   Furthermore, in order to have a sample that is consistent with the post-mergers (see next sub-section) we require that each galaxy in the pair have a stellar mass of log ($M_{\star}/M_{\odot}) \ge 10$.   There are 21,235 such galaxies that fulfill all of these criteria and that represent our pairs sample\footnote{Strictly, this is a sample of galaxies with close companions, rather than a true sample of pairs, because the closest companions identified do not always find unique pairings.  I.e. Galaxy A may have galaxy B as its closest companion, but galaxy B's closest neighbour may be galaxy C.  This explains why we have an odd number of galaxies in the `pairs' sample.  Nonetheless, we use the nomenclature `pairs' as it is widely adopted in the literature.}.  The pairs sample is dominated by major mergers, with two thirds of the sample exhibiting a mass ratio within 2:1.

\subsection{The post-merger sample}\label{pm_sec}

The post-merger sample is identified by applying the \textsc{mummi} neural network framework to the Ultraviolet Near Infrared and Optical Northern Survey (UNIONS).   Since \textsc{mummi} has been extensively described in Ferreira et al. (2024) and Ferreira et al. (in prep), we give only the briefest of details here.

\smallskip

\textsc{mummi} is a machine vision based merger classifier that uses an ensemble of 10 vision transformers and 10 convolutional neural networks to produce a total of 20 `votes'.  The classifications use all information in the images including streams, shells and bridges, without explicit (human) guidance on what to look for.  Highly pure (but incomplete) merger samples can be assembled by requiring a unanimous outcome in which 20/20 networks assign a positive merger classification.  However, Ferreira et al. (2024) show that a majority vote ($>$10/20) represents an excellent compromise between purity and completeness, and is hence the threshold adopted in our previous and current works.

\smallskip

\textsc{mummi} was trained on galaxies identified in the IllustrisTNG 100-1 (hereafter, TNG) simulation (Nelson et al. 2019).  The TNG training sample consists of a balanced set of mergers and non-interacting galaxies (no merger with a mass ratio $>$1:10 within the last 2 Gyr) with log(M$_{\star}$/M$_{\odot}$) $>$ 10 at $z<1$.   In addition to labelling the training sets as either non-mergers, pairs (that will merge within the next 1.7 Gyr) or post-mergers, the latter category is also labelled with the number of simulation snapshots since coalescence, where each snapshot is separated by $\sim$160 Myr on average.

\smallskip

Since we ultimately wish to apply the merger classifier to actual survey data obtained by UNIONS, it is important to train \textsc{mummi} with realistic data (Bottrell et al. 2019; Bickley et al. 2024b).     Mock $r$-band images of the TNG-selected galaxies are therefore generated with the same image qualities as UNIONS observations (including insertion into real frames to replicate CCD defects and foreground/background objects).  With this combined set of information, \textsc{mummi} is able to predict not only the merger status of a galaxy (non-merger, pair or post-merger) but also the time since coalescence ($T_{PM}$) in one of four bins, consolidated for maximal performance: $0<T_{PM}<0.16$, $0.16<T_{PM}<0.48$, $0.48<T_{PM}<0.96$ and $0.96<T_{PM}<1.76$ Gyr (see Ferreira et al. in prep).  The increasing length of the time bins reflects the increasing difficulty encountered by \textsc{mummi} to make accurate time predictions as the merger progresses.

\smallskip

Note that, in the work presented here, we take $T_{PM}$ = 0 to be the first snapshot after coalescence has occcured, although in practice the actual merger will have happened at some point in the preceding snapshot interval (of $\sim$ 160 Myr duration).  After imposing a threshold for prediction accuracy, post-merger galaxies are correctly assigned into these $T_{PM}$ bins 70--80 per cent of the time (Ferreira et al., in prep).  This threshold is based on the quality of the probability distributions produced by \textsc{mummi}, and removes any uncertain or spurious classifications mitigating any performance degradation between simulationa and observational domains (Ferreira et al. in prep).  Moreover, we found that \textsc{mummi}'s performance was constant across all mass ratios used in the training (from 1:1 down to 1:10).

\smallskip

Once trained, \textsc{mummi} is applied to the UNIONS fifth data release (DR5) of $r$-band imaging.  In addition to being the deepest available imaging, using the single band $r$ images means our merger identification is not biased by interaction-induced effects such as starbursts.  Using the $r$-band images there ensures both a sensitive selection and one driven purely by morphology.

\smallskip

Although DR5 covers a total of just under 5000 deg$^2$, we only apply \textsc{mummi} to the overlap with the SDSS DR7, thus providing complementary spectroscopic information for the post-merger sample.  Ferreira et al. (2024) identify 235,354 galaxies with $0.03 < z < 0.3$ in this overlap region and present catalogs of galaxy mergers thus identified.  Since the \textsc{mummi} pipeline was trained using galaxies whose stellar mass log($M_{\star}/M_{\odot}) > 10$, we impose the same limit on the observational sample.  Our final sample, for which we have both SDSS DR7 spectroscopy and \textsc{mummi} merger predictions, is 231,340 galaxies of which 11,309 are identified as post-mergers under the majority voting scheme.  Imposing the recommended quality threshold for time bin predictions (see Ferreira et al. 2024; Ferreira et al. in prep) reduces this to 8141 post-merger galaxies of which there are 785, 740, 778 and 5838 galaxies in the $0 < T_{PM} <$ 0.16 Gyr, $0.16 < T_{PM} <$ 0.48 Gyr, $0.48 < T_{PM} <$ 0.96 Gyr and $0.96 < T_{PM} <$ 1.76 Gyr time bins, respectively.  The larger sample in the last time bin reflects both the longer time interval assigned to this bin (0.8 Gyr), compared to the shorter $T_{PM}$ bins, as well as relative levels of completeness (more details on \textsc{mummi}'s performance are provided in Ferreira et al. in prep).

\subsection{Control matching}\label{control_sec}

\begin{figure}
	\includegraphics[width=8.5cm]{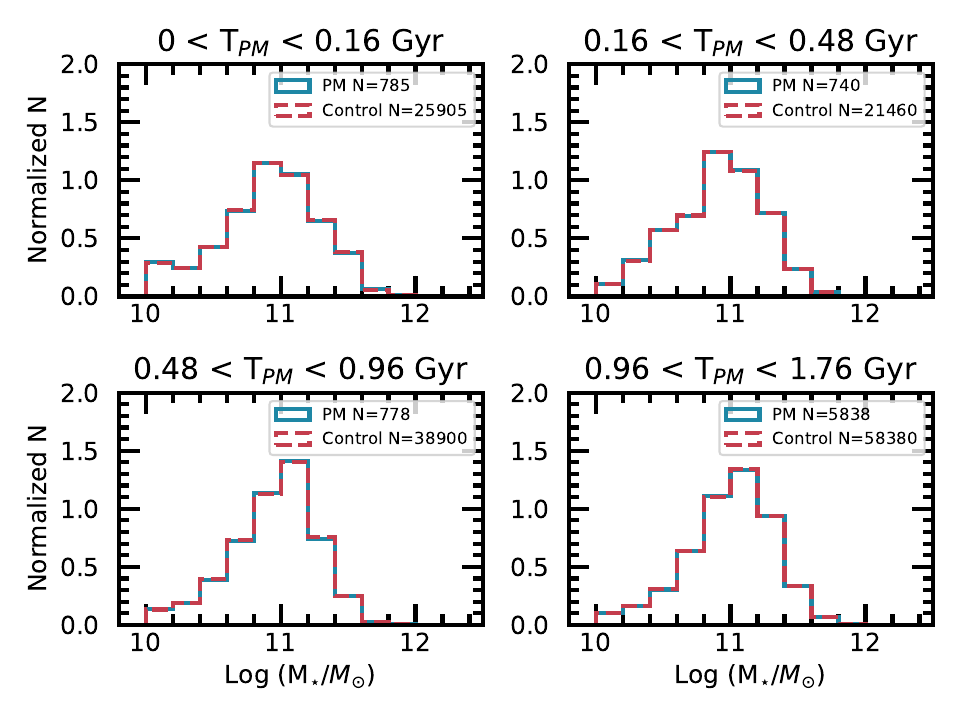}
        \caption{The stellar mass distributions of the post-merger sample in the four $T_{PM}$ time bins (blue) and matched control sample (red dashed). }
        \label{mhist}
\end{figure}

\begin{figure}
	\includegraphics[width=8.5cm]{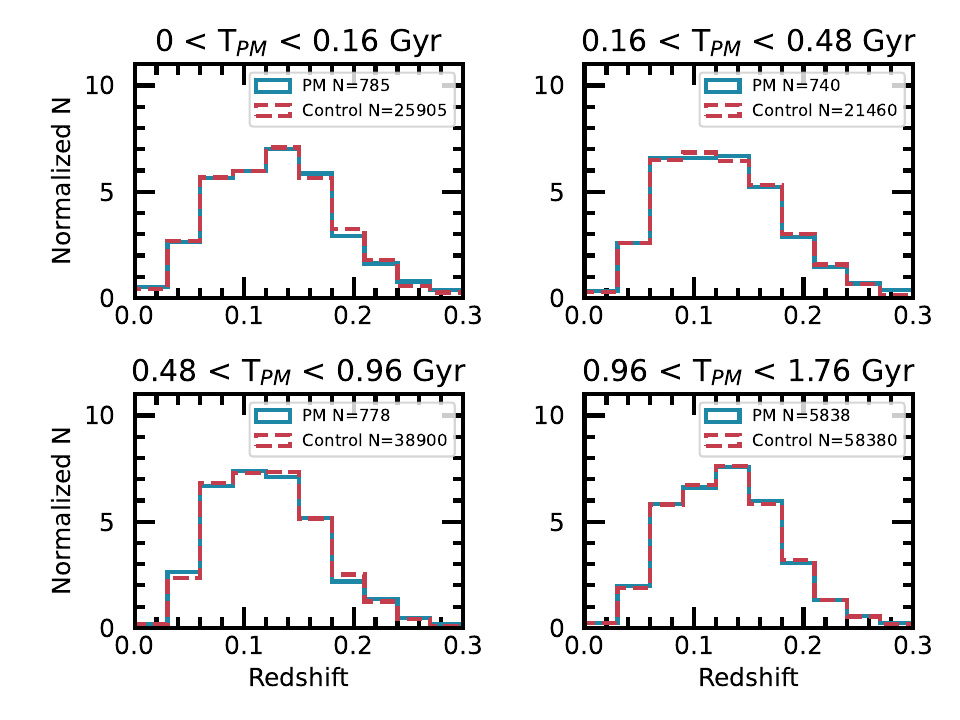}
        \caption{The redshift distributions of the post-merger sample in the four $T_{PM}$ time bins (blue) and matched control sample (red dashed). }
        \label{zhist}
\end{figure}

A control sample of non-merger galaxies is constructed for both the pairs and the post-merger samples, in order that their AGN fractions can be differentially assessed.  The matching procedure is run separately for the pairs and for each of the four $T_{PM}$ samples, but the same over-arching methodology is applied to all samples.

\smallskip

First, a pool of potential control galaxies is identified for both the pair and post-merger samples.  The galaxies in these control pools are essentially all galaxies that do not qualify as either pairs or post-mergers.  Specifically, for the pairs, the control pool must represent galaxies without a close companion, so we require $r_p > 100$ kpc.   We also remove from the control pool galaxies identified as post-mergers in the Galaxy Zoo sample of Ellison et al. (2013).  There are 508,143 galaxies in the SDSS DR7 that fulfill these criteria and hence constitute the control pool for the pairs.

\smallskip

For the post-mergers, the control pool consists of galaxies with no more than 2/20 of the neural networks assigning a merger classification ($p>0.5$), yielding 160,009 galaxies for this control pool.  Modifications to these criteria were investigated and found not to yield significant changes to our results.

\smallskip

The control matching follows the same procedure as laid out in our previous papers in this series.  An iterative approach finds the best simultaneous match in both stellar mass and redshift (with equal weighting) for all mergers in a given sample (e.g. the pairs) until a Kolmogorov-Smirnov (KS) test returns a value of $p<0.99$ or a maximum of 50 controls per merger are found.   In this way, the mergers in each sample (i.e. the pairs and post-mergers in each of the four time bins) have the same number of matched controls, but that number can be different from one sample to the next.  Also, since the matching is run separately for the pairs and the four post-merger samples, whilst a given control can not be re-used in a given sample, it can be re-used for different samples (e.g. pairs and four post-merger time bins).  The number of controls for each galaxy in the four time bins is (from shortest to longest $T_{PM}$) 33, 29, 50 and 10, where the smaller number of controls in the longest $T_{PM}$ bin is driven by the much larger sample ($\sim 5800$ compared with $\sim$ 800 in shorter time bins).  The pairs sample achieves 9 matches per galaxy; again, the smaller number of control galaxies is driven by the larger size of the pairs sample ($\sim$ 21,000).

\smallskip

As a demonstration of the excellent quality of the matching, we show in Figures \ref{mhist} and \ref{zhist} the normalized histogram distributions of stellar mass and redshift for the four post-merger time bins, where the post-merger distributions are shown in blue and the controls in red.  The legend reports the total number of controls identified for each sample. It can be seen that both the stellar mass and redshift distributions are indistinguishable between mergers and controls.  Moreover, the properties of the post-mergers in the different time bins are seen to be very consistent.  For example, the median log stellar mass (in units of solar masses) from the shortest to longest time bins is 10.95, 10.93, 10.99, 11.03.

\subsection{Classification of AGN}\label{agn_sec}

\begin{table*}
  \begin{center}
  \caption{Statistics for AGN identification using any of the three metrics (NLAGN, BLAGN and mid-IR colour) used in this paper.  In each time bin, we tabulate the number of post-mergers, number of AGN in the post-merger sample, the fraction of post-mergers that are AGN and then the equivalent statistics for the control sample.  The final column shows the ratio of the fraction of AGN in the post-mergers and controls, hence representing an AGN excess triggered by the interaction.}
\begin{tabular}{lccccccc}
\hline

$T_{PM}$ (Gyr) & $N_{PM}$ & $N_{PM,AGN}$ & $f_{PM,AGN}$ & $N_{control}$ & $N_{control,AGN}$ & $f_{control,AGN}$ & Excess \\ 

\hline                                                                                                      
0-0.16    &  785 & 168 & 0.21$\pm$0.01   & 25905 & 1789 & 0.069$\pm$0.0001 & 3.1$\pm$0.2 \\
0.16-0.48 &  740 & 104 & 0.14$\pm$0.01   & 21460 & 1504 & 0.070$\pm$0.0002 & 2.0$\pm$0.2\\
0.48-0.96 &  778 &  69 & 0.09$\pm$0.01   & 38900 & 2293 & 0.059$\pm$0.0001 & 1.5$\pm$0.2\\
0.96-1.76 & 5838 & 302 & 0.052$\pm$0.002 & 58380 & 3374 & 0.058$\pm$0.0001 & 0.9$\pm$0.1\\
\hline
\end{tabular}
\label{agn_all_tab}
  \end{center}
\end{table*}

AGN are a multi-wavelength phemomenon whose various structures can be detected across the electromagnetic spectrum.  The detectability of an AGN at a given wavelength depends on a range of factors including viewing angle, obscuration and bolometric dominance (see reviews by Ramos Almeida \& Ricci 2017; Hickox \& Alexander 2018).  Some studies of AGN triggering in mergers have been (often, necessarily or deliberately) limited to single AGN diagnostics, such as the flux ratios of narrow emission lines (e.g. Li et al., 2008; Ellison et al. 2008, 2011, 2013), X-rays (e.g. Koss et al. 2010, 2018) or mid-IR colours (e.g. Goulding et al. 2018; Barrows et al. 2023).  However, since not all AGN are `visible' at all wavelengths, a complete census requires the combination of multiple identification techniques.  Moreover, by comparing the incidence of AGN using different techniques it is possible to infer physical properties of the accretion process, such as the prevalence of dust, or the mode of accretion.  Thanks to the availability of large area (or even all sky) surveys over a range of wavelengths, it has therefore become more common to assess simultaneously the incidence of AGN identified in the optical, mid-IR and radio (e.g. Satyapal et al. 2014; Ellison et al. 2015, 2019; Weston et al. 2017; Gao et al. 2020).  Finally, the identification of AGN using X-rays is highly complementary to other identification techniques, providing further clues about the extent of obscuration in interacting galaxies (e.g. Secrest et al. 2020; Li et al. 2023; Comerford et al. 2024; Dougherty et al. 2024).  In this regard, the deep and wide area coverage of eROSITA has proven an excellent companion to other large surveys for the assessment of the merger AGN connection (e.g. Bickley et al. 2024a; La Marca et al. 2024).

\smallskip

Unfortunately, the overlap between the German consortium (public) eROSITA coverage and UNIONS is minimal, precluding an assessment of Xray AGN in the current work.   In the work presented here we thus limit ourselves to just three complementary AGN diagnostics.  First, we identify narrow line AGN (NLAGN) using line flux ratios of [OIII]$\lambda$5007/H$\beta$ and [NII]$\lambda$6583/H$\alpha$, taken from the MPA/JHU catalog.    The selection of NLAGN identifies galaxies with an extended (typically within $\sim$1 kpc of the nucleus, but sometimes on larger scales, e.g. Hainline et al. 2013) region of gas that is excited by radiation from the AGN.  Due to its extent, narrow lines are generally not obscured by the torus.  Various demarcations have been proposed to distinguish AGN from star forming regions in the flux ratio diagrams (e.g. see Kewley et al. 2019 for a review).  In the work presented here we adopt the stringent cut defined by Kewley et al. (2001) and require a S/N in all four lines of at least 3. We repeated all of the experiments presented in this paper using different S/N cuts from 2 to 5, and also with the more lenient Kauffmann et al. (2003b) cut and found that these choices did not qualitatively affect our results.

\smallskip

Second, we use mid-IR colours that identify emission from the dusty torus on a scale of $\sim$ 10 pc from the nucleus.  In the work presented here, we use the forced photometry measurements in the unWISE catalog (Lang et al. 2016), specifically making use of magnitudes in the W1 (3.4 $\mu$m) and W2 (4.6 $\mu$m) bands, whose combination is known to be an effective identifier of AGN (Stern et al. 2012; Assef et al. 2013).  Specifically, we adopt a W1$-$W2$>0.5$ (in Vega magnitudes) cut, which was shown by Blecha et al. (2018) to yield a fairly complete AGN census at low redshift.  However, since the relative contribution of AGN and stellar dust heating can affect mid-IR AGN selection (e.g. Figure of Donley et al. 2012) we will explicitly investigate the impact of the W1$-$W2 threshold below.

\smallskip

Finally, we identify broad line AGN (BLAGN), whose emission emanates from within the dusty torus on scales of $\sim$ 1 pc.  The identification of BLAGN is therefore expected to be sensitive to both the viewing angle and the amount/distribution of nuclear dust.   In the work presented here, we use the catalog produced by Liu et al. (2019), in which BLAGN are identified if the fit to the emission lines is significantly improved by the presence of a second, broadened component.  The minimum velocity width of the broad components fit by Liu et al. (2019) is $\sim$ 500 km/s, although the majority are in excess of 1000 km/s.  If a galaxy is identified as both a NLAGN and BLAGN, we follow Bickley et al. (2024a) and give preference to the latter classification, such that a galaxy only appears in one of the two optical emission line categories.

\begin{table*}
  \begin{center}
  \caption{Statistics for NLAGN identification.  In each time bin, we tabulate the number of post-mergers, number of AGN in the post-merger sample, the fraction of post-mergers that are AGN and then the equivalent statistics for the control sample.  The final column shows the ratio of the fraction of AGN in the post-mergers and controls, hence representing an AGN excess triggered by the interaction.}
\begin{tabular}{lccccccc}
\hline

$T_{PM}$ (Gyr) & $N_{PM}$ & $N_{PM,AGN}$ & $f_{PM,AGN}$ & $N_{control}$ & $N_{control,AGN}$ & $f_{control,AGN}$ & Excess \\ 

\hline                                                                                                      
0-0.16    &  785 &  85 & 0.11$\pm$0.01     & 25905 & 926  & 0.036$\pm$0.001   & 3.0$\pm$0.3 \\
0.16-0.48 &  740 &  50 & 0.068$\pm$0.009   & 21460 & 760  & 0.035$\pm$0.001   & 1.9$\pm$0.3\\
0.48-0.96 &  778 &  34 & 0.044$\pm$0.007   & 38900 & 1443 & 0.037$\pm$0.001   & 1.2$\pm$0.2\\
0.96-1.76 & 5838 & 166 & 0.028$\pm$0.002   & 58380 & 2078 & 0.036$\pm$0.001   & 0.80$\pm$0.06\\
\hline
\end{tabular}
\label{nlagn_tab}
  \end{center}
\end{table*}

\begin{table*}
  \begin{center}
  \caption{Statistics for mid-IR AGN identification.  In each time bin, we tabulate the number of post-mergers, number of AGN in the post-merger sample, the fraction of post-mergers that are AGN and then the equivalent statistics for the control sample.  The final column shows the ratio of the fraction of AGN in the post-mergers and controls, hence representing an AGN excess triggered by the interaction.}
\begin{tabular}{lccccccc}
\hline

$T_{PM}$ (Gyr) & $N_{PM}$ & $N_{PM,AGN}$ & $f_{PM,AGN}$ & $N_{control}$ & $N_{control,AGN}$ & $f_{control,AGN}$ & Excess \\ 

\hline                                                                                                      
0-0.16    &  785 & 104 & 0.13$\pm$0.1      & 25905 & 858  & 0.033$\pm$0.001   & 4.0$\pm$0.4 \\
0.16-0.48 &  740 &  65 & 0.09$\pm$0.01     & 21460 & 734  & 0.034$\pm$0.001   & 2.6$\pm$0.3\\
0.48-0.96 &  778 &  50 & 0.051$\pm$0.008   & 38900 & 842  & 0.022$\pm$0.001   & 2.4$\pm$0.4\\
0.96-1.76 & 5838 & 161 & 0.028$\pm$0.002   & 58380 & 1237 & 0.021$\pm$0.001   & 1.3$\pm$0.1\\
\hline
\end{tabular}
\label{iragn_tab}
  \end{center}
\end{table*}

\begin{table*}
  \begin{center}
  \caption{Statistics for BLAGN identification.  In each time bin, we tabulate the number of post-mergers, number of AGN in the post-merger sample, the fraction of post-mergers that are AGN and then the equivalent statistics for the control sample.  The final column shows the ratio of the fraction of AGN in the post-mergers and controls, hence representing an AGN excess triggered by the interaction.}
\begin{tabular}{lccccccc}
\hline

$T_{PM}$ (Gyr) & $N_{PM}$ & $N_{PM,AGN}$ & $f_{PM,AGN}$ & $N_{control}$ & $N_{control,AGN}$ & $f_{control,AGN}$ & Excess \\ 

\hline                                                                                                      
0-0.16    &  785 &  33 & 0.042$\pm$0.007   & 25905 & 435 & 0.017$\pm$0.007   & 2.5$\pm$0.4 \\
0.16-0.48 &  740 &  30 & 0.041$\pm$0.007   & 21460 & 363 & 0.017$\pm$0.001   & 2.4$\pm$0.5\\
0.48-0.96 &  778 &  20 & 0.025$\pm$0.006   & 38900 & 456 & 0.0117$\pm$0.0005 & 2.2$\pm$0.4\\
0.96-1.76 & 5838 & 104 & 0.018$\pm$0.002   & 58380 & 721 & 0.0123$\pm$0.0005 & 1.4$\pm$0.2\\
\hline
\end{tabular}
\label{blagn_tab}
  \end{center}
\end{table*}

\section{Results}\label{results_sec}

\subsection{An overall census of AGN triggering}\label{all_agn_sec}

\begin{figure}
	\includegraphics[width=8.5cm]{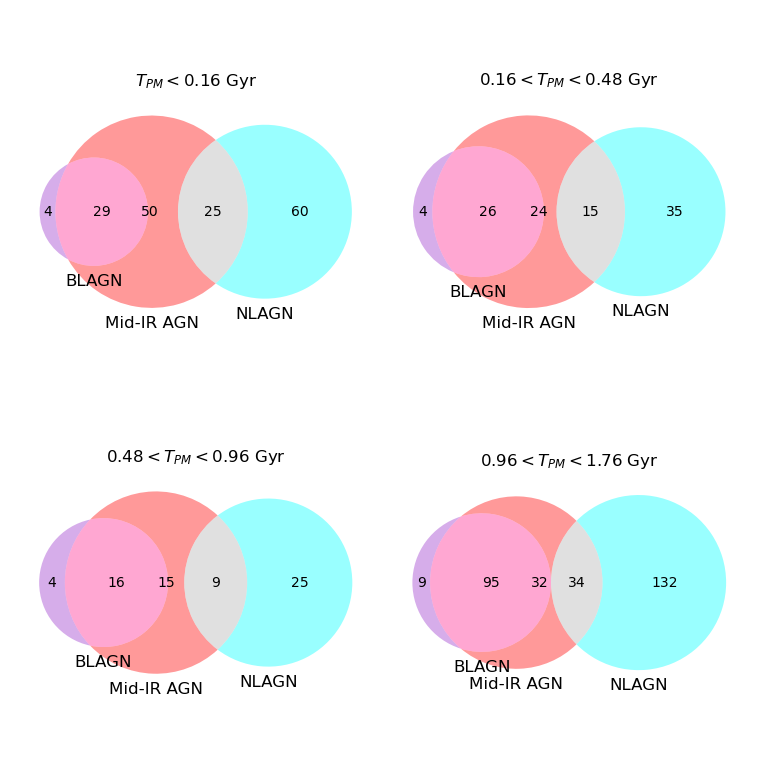}
        \caption{The number statistics of AGN in the post-merger sample represented as a Venn diagram for each of the four $T_{PM}$ bins.  The diagram is schematic only (areas of the circles are not to scale).  The values in each diagram represent, from left to right, the number of galaxies that are only BLAGN, the number of galaxies that are both BLAGN and mid-IR AGN, the number of galaxies that are only mid-IR AGN, the number of galaxies that are mid-IR AGN and NLAGN and the number of galaxies that are only NLAGN.  NLAGN and BLAGN are mutually exclusive by construction.  }
        \label{venn_fig}
\end{figure}

Figure \ref{venn_fig} presents the overall incidence of AGN in our post-merger sample as a Venn diagram.  A separate Venn diagram is presented for each of the $T_{PM}$ bins.  Although there is overlap between AGN samples selected using different diagnostics (although recall that the NLAGN and BLAGN classes are mutually exclusive by design), some sources are uniquely identified using just one metric.  In order to attempt the most complete census of AGN in our merger and control samples we begin by counting the number of galaxies that are identified by any of the three diagnostics used in this work.  In this way, we are including AGN with a range of luminosities and dust obscuration properties and quantifying an overall enhancement in merger driven triggering.  This is a useful first step before considering the three AGN diagnostics separately, which in turn allows us to understand the evolving AGN demographics.

\smallskip

In Table \ref{agn_all_tab} we compile the statistics for merger and control samples in each of the four $T_{PM}$ bins (since pair statistics have been published in many of our previous works, we focus here on the novel post-merger sample).  We tabulate the number of post-mergers and controls in each time bin and both the number and fraction of galaxies in each of these samples identified as AGN by at least one of the three diagnostics presented in Section \ref{agn_sec}.  Finally, we compute an AGN excess, which is defined as the fraction of mergers (in a given $T_{PM}$ bin) with an AGN, divided by the fraction of control galaxies (in the same $T_{PM}$ bin) with AGN.  Thus the AGN excess is equal to one if the incidence of AGN in mergers is consistent with expectations given the distribution of stellar mass and redshift.  Figure \ref{agn_all_fig} presents the AGN excesses reported in Table \ref{agn_all_tab} graphically, and also includes the statistics for the pairs sample, where the excess is computed in each $r_p$ bin.  The figure can be read from left to right, starting with widely separated pairs, approaching one another as the interaction progresses, and then onwards in units of time after coalescence.

\smallskip

Figure \ref{agn_all_fig} demonstrates a clear temporal evolution in prevalence of AGN during the merger sequence.  At wide separations ($r_p>60$ kpc), the incidence of AGN in galaxy pairs is statistically consistent with the control sample.  At smaller separations, we begin to see an enhancement in the frequency of AGN, with an excess of a factor of $\sim$ 2 in the closest pairs.  Qualitatively similar trends with pair separation have been seen in numerous previous works, although the level of enhancement and $r_p$ extent out to which an excess is seen depends on selection method (e.g. Ellison et al. 2011; Satyapal et al. 2014; Steffen et al. 2023).

\smallskip

The main novelty of the current study is our ability, for the first time, to trace the excess of AGN into the post-coalescence regime as a function of $T_{PM}$.  The right hand side of Figure \ref{agn_all_fig} shows that the peak in AGN excess happens shortly after coalescence, with the AGN fraction increasing from 7 per cent in the control sample to 21 per cent in post-mergers (Table \ref{agn_all_tab}), leading to a factor of $\sim$3 enhancement in the $0 < T_{PM} < 0.16$ Gyr bin.  As the post-merger remnant ages, the AGN excess declines, with no statistical enhancement beyond $T_{PM}$ = 1 Gyr.  We emphasize that although statistical AGN excesses are seen over the majority of the merger time sequence (from a pair separation of 60 kpc to a post-merger timescale of $\sim$ 1 Gyr), this does not mean that a single AGN event lasts for this entire duration.  Instead, our results are consistent either with accretion events being longer lived in mergers, or more frequent, or both, but we can place no constraints on the actual duration or duty cycle.

\subsection{AGN excess by diagnostic}

\begin{figure}
	\includegraphics[width=8.5cm]{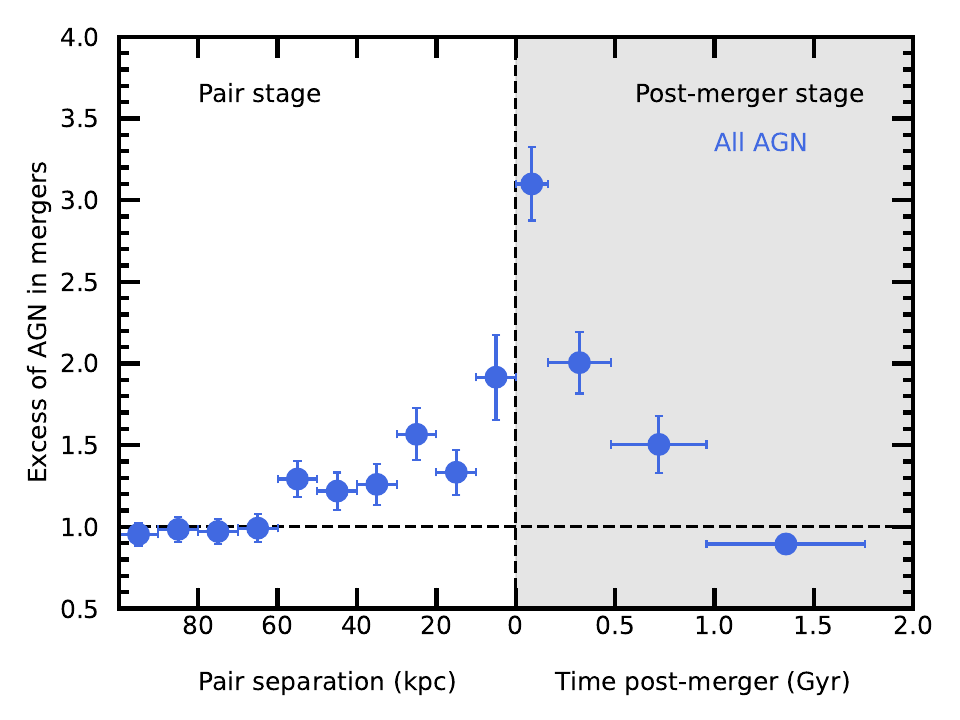}
        \caption{The excess of AGN host galaxies identified by any of the diagnostics used in this paper (NLAGN, BLAGN and mid-IR colours) in the pairs (left side of the diagram) and post-mergers (right side of the diagram) compared with the control sample. The horizontal dashed line shows an excess of one, i.e. the same AGN fraction in mergers and controls.   The pair sample is binned in units of projected separation (kpc) whereas the post-mergers are binned by the time post-merger, with x- and y-axis error bars indicating the width of the $r_p$ or $T_{PM}$ bin and binomial uncertainties respectively.}
        \label{agn_all_fig}
\end{figure}

\begin{figure*}
	\includegraphics[width=6cm]{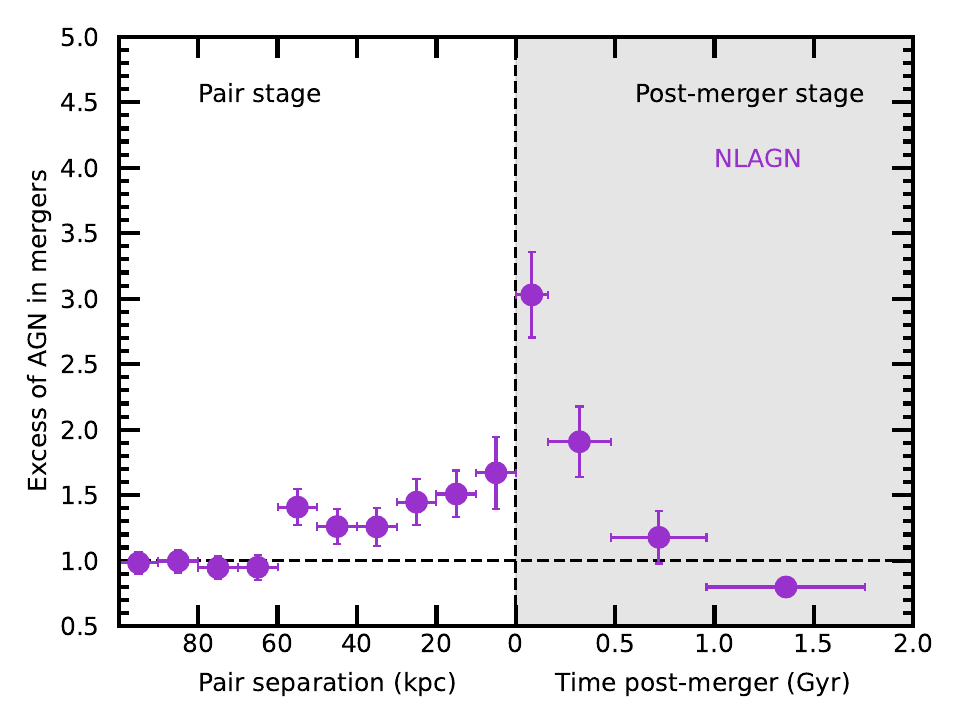}
	\includegraphics[width=6cm]{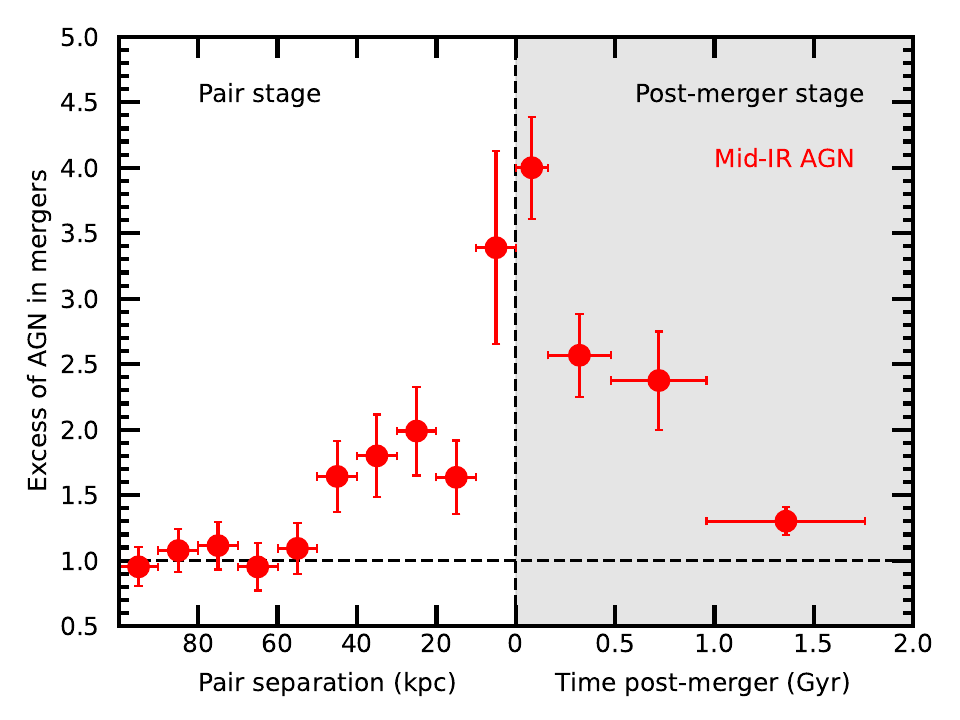}
	\includegraphics[width=6cm]{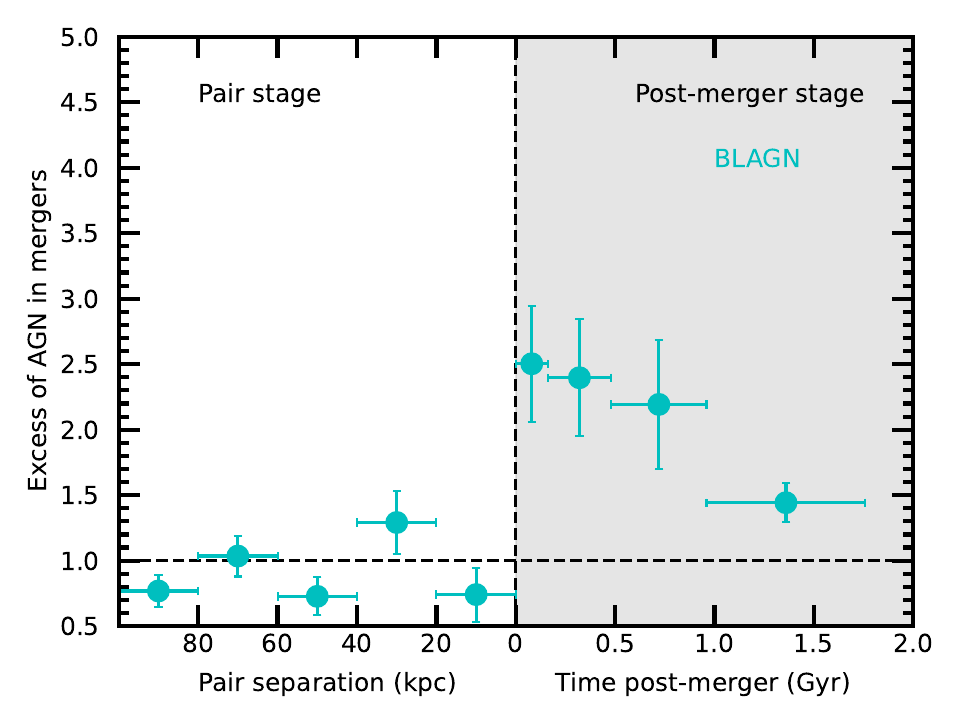}
        \caption{The excess of AGN host galaxies identified by each individual diagnostic in the pairs (left side of the diagram) and post-mergers (right side of the diagram) compared with the control sample. From left to right the panels show the excess of NLAGN, mid-IR selected AGN and BLAGN.  The y-axis extent is fixed for all panels in order to facilitate comparison. }
        \label{agn_3}
\end{figure*}

The results presented in Figure \ref{agn_all_fig} demonstrate that, when all diagnostics are considered together, AGN triggering begins in the pair phase, peaks around the time of coalescence and lasts until $\sim$ 1 Gyr post-merger.  We next evaluate the AGN frequencies separately for each of the three AGN diagnostics, whose comparative excesses can inform us about levels of dust obscuration and nuclear structure.  Tables \ref{nlagn_tab} to \ref{blagn_tab} compile the statistics for NLAGN, mid-IR selected AGN and BLAGN in the four post-merger time bins, as well as their respective control samples.  AGN excesses are tabulated in the final columns.  The AGN excesses are also presented graphically in Figure \ref{agn_3}, showing (from left to right) the enhancement in AGN incidence in NLAGN, mid-IR AGN and BLAGN. All three panels have the same y-axis to facilitate comparison of the statistics between AGN diagnostics.

\smallskip

Starting with the left hand panel of  Figure \ref{agn_3}, we find that the qualitative shape of the NLAGN excess closely matches that of our overall AGN census in Figure \ref{agn_all_fig}, including a peak in the AGN excess at coalescence and the deficit of AGN in the last $T_{PM}$ bin.  The excess of mid-IR AGN (middle panel of Figure \ref{agn_3}) also shows the same overall temporal evolution, with a positive signal arising for pairs within $r_p<60$ kpc and peaking around the time of coalescence.  However, the levels of enhancement are somewhat larger in the mid-IR selected AGN ($\times$3 in the closest pairs and $\times$4 at coalescence) compared with NLAGN ($\times$1.5 in the close pairs and $\times$3 at coalescence).  Another important difference between the NLAGN and mid-IR statistics is that whereas the former showed no excess beyond $\sim$1 Gyr, mid-IR AGN show an elevated frequency in post-mergers out to at least 1.76 Gyr.

\smallskip

Finally, in the right hand panel of Figure \ref{agn_3} we show the statistics for BLAGN, where a quite different behaviour is observed.  In contrast to the NLAGN and mid-IR selected AGN, the BLAGN show no excess in the pair phase, and in some of the bins there is even a marginal deficit\footnote{Due to the smaller numbers of BLAGN compared with other diagnostics, 20 kpc bins are used in the right hand panel of Figure \ref{agn_3} compared with 10 kpc bins for NLAGN and mid-IR.}.  However, after coalescence the BLAGN quickly become more frequent with an excess that persists out to the longest time bin in our sample (1.76 Gyr), just as we saw in the case of mid-IR AGN.

\smallskip

Taken together, the three panels in Figure \ref{agn_3} demonstrate not only that mergers trigger AGN, but that the statistically enhanced level of AGN activity is observed over a long time period in the merger sequence, beginning when the galaxies are still well separated and persisting until at least $\sim$ 1.8 Gyr post-coalescence, depending on AGN diagnostic.  However, the different AGN diagnostics behave qualitatively differently.  As discussed in Section \ref{agn_sec}, the narrow lines, broad lines and mid-IR emission originate from distinct physical structures within the central few kpc of the galaxy.  The different behaviours seen in Figure \ref{agn_3} therefore allow us to develop a picture of how the nuclear region itself is changing throughout the merger.  Specifically,  we can interpret the results of the three panels in Figure \ref{agn_3} within the context of dust obscuration and AGN luminosity; we explore further these demographics in the following sub-section.

\subsection{Obscuration through the merger sequence}

Several previous studies (e.g. Satyapal et al. 2014; Weston et al. 2017; Gao et al. 2020; Bickley et al. 2023) have seen elevated frequencies of mid-IR selected AGN compared with NLAGN and suggested that this is caused by enhanced nuclear obscuration during a merger.  The addition of BLAGN statistics both supports and adds to this narrative.  On the one hand, the enhancement of NLAGN and mid-IR selected AGN in the pair regime shows that nuclear accretion is triggered in this early encounter phase.  However, the lack of a concomitant enhancement in BLAGN (and possibly even a small deficit) is consistent with higher covering fractions of nuclear obscuring material in the early stages of the interaction (see also Bickley et al. 2024a), when gas and dust are expected to accumulate in the central regions of the galaxy (Blecha et al. 2018).  The emergence of a BLAGN excess after coalescence thus requires not only that AGN are more frequent in this regime (as already found using mid-IR and NLAGN diagnostics), but also that the broad line region is once again visible.  Our results are therefore consistent with a picture in which the covering fraction of gas and dust is subsequently reduced promptly after coalescence, in agreement with Blecha et al. (2018) who found a strong reduction in line of sight $N_H$ only 40 Myr after the merger.  It is unlikely that the dust is completely removed (as in the classical `blowout' picture) since nuclear dust is still required in order to see the mid-IR AGN excess seen through post-coalescence phase in the central panel of Figure \ref{agn_3}.  The results in Figure \ref{agn_3} are therefore consistent with mergers triggering AGN in combination with an evolution in the unobscured AGN fraction (or, equivalently, evolution in the dust covering fraction) during the merger sequence.

\smallskip

\begin{figure}
	\includegraphics[width=8.5cm]{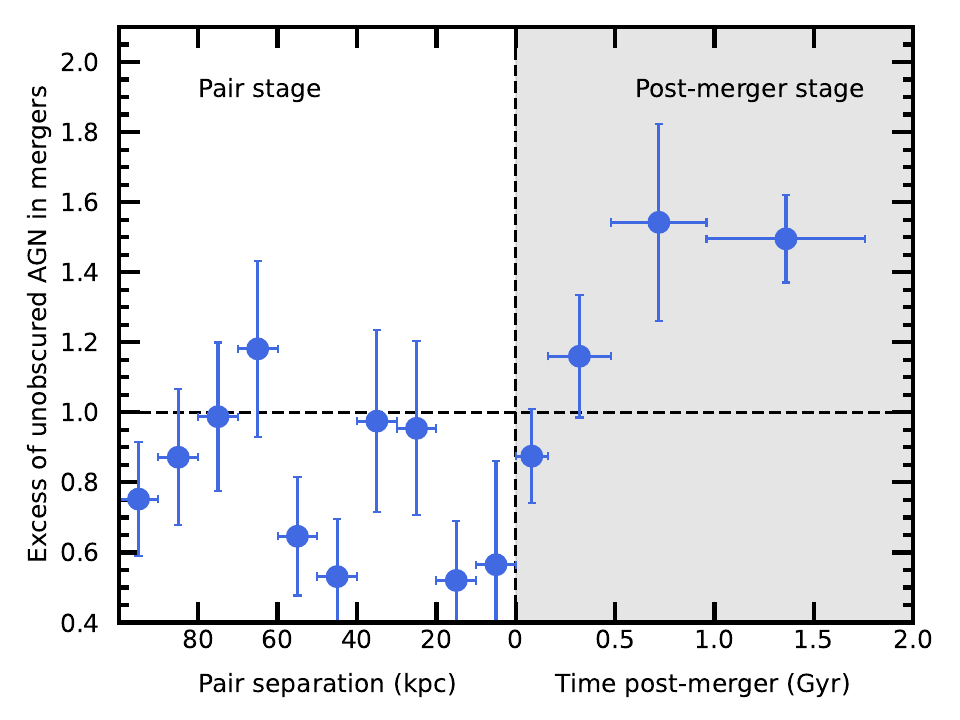}
        \caption{The excess of unobscured emission line AGN as a function of merger stage.  The unobscured fraction of both merger and control samples is computed as $N_{BLAGN} / (N_{BLAGN} + N_{NLAGN})$, with the excess representing the ratio of fractions in the mergers relative to the controls.  Although of marginal significance, the low values in the pair regime hint that obscuration may be enhanced pre-coalescence.  Conversely, in the post-merger regime the unobscured AGN become progressively more common, consistent with feedback impacting the central dust distributions. }
        \label{unobs_agn}
\end{figure}

An increase in the level of dust obscuration in galaxies is often assessed by changes in the Balmer decrement.  However, we see no difference in the ratio of the H$\alpha$/H$\beta$ line fluxes in NLAGN as a function of merger stage, which is probably due to the more spatially extended location of the narrow line emitting gas compared with the nuclear dust we are trying to probe.  We therefore need an alternative method to assess the dust obscuration in the central region, by incorporating the broad line diagnostic.   Recalling that, in our sample, emission line AGN can only be classified as either BLAGN or NLAGN (not both), we define the fraction of unobscured emission line AGN as

\begin{equation}\label{eqn_unobs}
  f_{unobs} = N_{BLAGN} / (N_{BLAGN}  + N_{NLAGN}).
\end{equation}

\begin{figure*}
	\includegraphics[width=6cm]{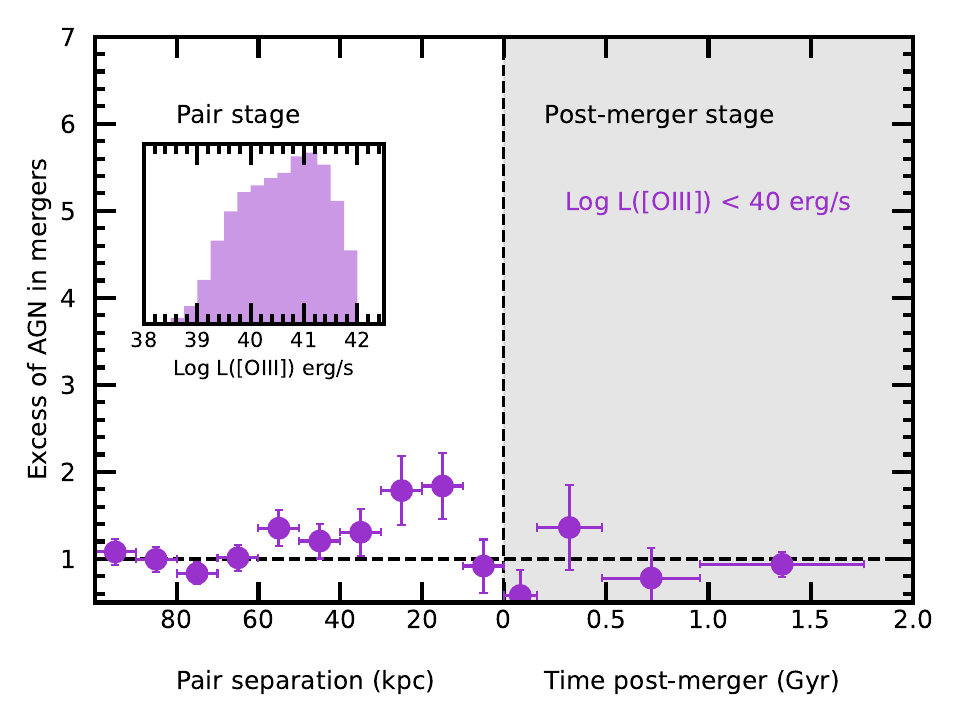}
	\includegraphics[width=6cm]{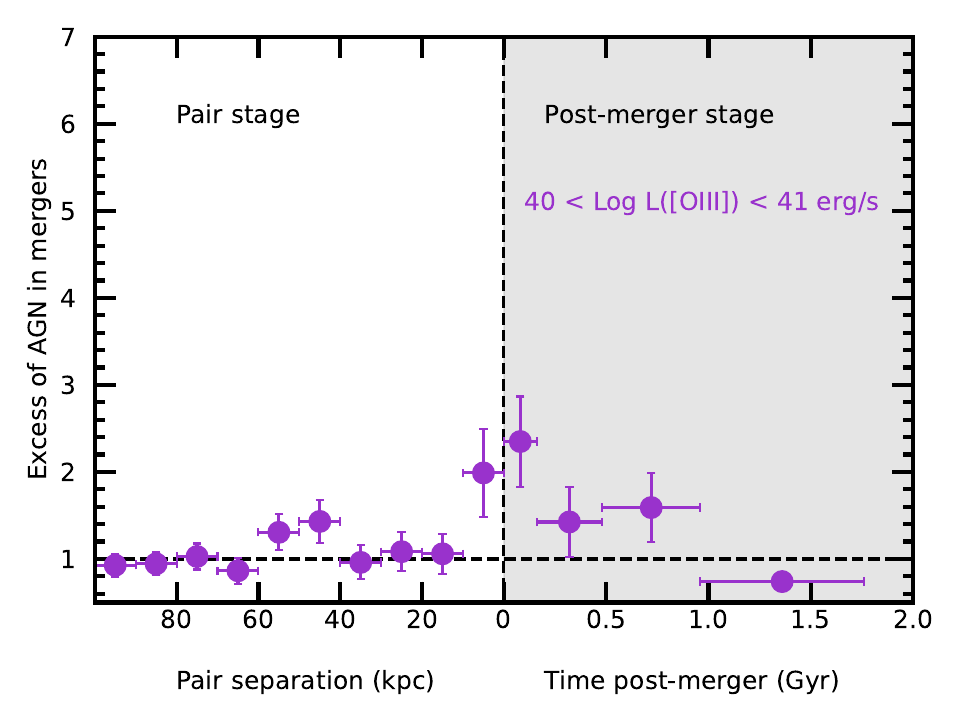}
	\includegraphics[width=6cm]{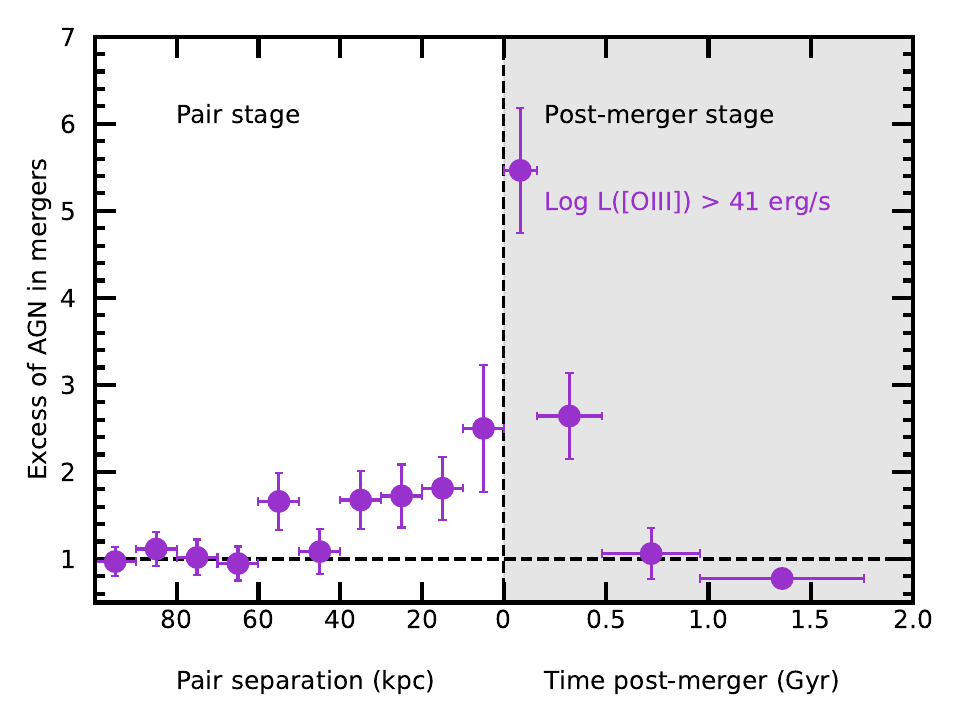}
        \caption{The excess of NLAGN host galaxies as a function of the [OIII]$\lambda$5007 line luminosity. Each panel counts NLAGN within a fixed range of L([OIII]).  From left to right the panels show the excess of low (log L([OIII]) $<$ 40 erg/s), medium (40 $<$ log L([OIII]) $<$ 41 erg/s) and high (log L([OIII]) $>$ 41 erg/s) luminosity AGN respectively.  The y-axis extent is fixed for all panels in order to facilitate comparison. The inset histogram shows the L([OIII]) distribution for the full SDSS NLAGN sample as a reference for the distribution of the sample.}
        \label{lo3_xs}
\end{figure*}

$f_{unobs}$ therefore represents the fraction of galaxies identified as an AGN through their emission lines in which the broad line region is visible.  The number of mid-IR AGN does not appear in Equation \ref{eqn_unobs} since some galaxies are classified as both mid-IR and emission line AGN (i.e. NLAGN or BLAGN, see Figure \ref{venn_fig}), hence their inclusion would lead to double counting.  We compute $f_{unobs}$ for both mergers (pairs and post-mergers) and controls in bins of $r_p$ and $T_{PM}$.  The ratio of  $f_{unobs}$ in the mergers compared with the controls is the unobscured AGN excess.  Figure \ref{unobs_agn} shows that in the pair phase, there is a deficit of unobscured AGN in several of the projected separation bins, consistent with our interpretation from Figure \ref{agn_3} (BLAGN excesses below one in the pre-coalescence regime).  As we move into the post-merger regime, the unobscured AGN fraction increases with $T_{PM}$ showing that late stage post-merger remnants have lower covering fractions of dust compared with AGN in galaxies in the control sample.  Figure \ref{unobs_agn} therefore supports our suggestion that Figure \ref{agn_3} can be interpreted as a combination of mergers triggering AGN, combined with an evolution of the nuclear dust structure.

\subsection{The luminosities of merger triggered AGN}

Despite the broad agreement in the literature that (at least at low redshift) mergers are capable of triggering AGN, there remains considerable debate about the luminosity dependence of this process.  Treister et al. (2012) used a compilation of literature data to show a strong positive correlation between the luminosity of AGN and the fraction associated with mergers.  Since then, there have been numerous papers both claiming (e.g. Marian et al. 2020; Pierce et al. 2022; La Marca et al. 2024) and refuting (e.g. Villforth et al. 2017; Villforth 2023) a link between the most luminous AGN and mergers at low redshift.  In simulations, there does seem to be a link between mergers and more luminous AGN (McAlpine et al. 2020; Byrne-Mamahit et al. 2023, 2024), although not all AGN can be associated with a merger, even at the highest luminosities or when very minor mass ratios are considered (Byrne-Mamahit et al. in prep).

\smallskip

Most of these previous works, both using observations and simulations, have assessed what fraction of AGN (as a function of luminosity) can be associated with a galaxy-galaxy interaction.  Such an approach requires starting with a sample of AGN and then identifying mergers.  Here, we can take a complementary approach, by starting with our merger sample and investigating the luminosities of the sample in comparison with the control sample.  In Figure \ref{lo3_xs} we once again compute the excess of NLAGN in the pairs and post-merger samples, but now only label NLAGN if they fall into a requisite luminosity range.  Specifically, we compute the luminosity of the [OIII]$\lambda$5007 line (hereafter, simply L([OIII])), which, when the galaxy is AGN dominated (as we expect it to be for our NLAGN as they are above the Kewley et al. 2001 demarcation) is indicative of AGN power. The left, middle and right hand panels of Figure \ref{lo3_xs} show the AGN excess for low (log L([OIII]) $<$ 40 erg/s), medium (40 $<$ log L([OIII]) $<$ 41 erg/s) and high (log L([OIII]) $>$ 41 erg/s) luminosity AGN.  For reference, the overall distribution of log L([OIII]) of NLAGN in the full SDSS sample is shown as a histogram inset in the first panel.  From this inset histogram, it can be seen that the high luminosity bin (log L([OIII]) $>$ 41 erg/s) represents the most powerful NLAGN found in SDSS.  

\smallskip

Figure \ref{lo3_xs} demonstrates that the enhancement in the incidence of NLAGN in post-mergers shown in Figure \ref{agn_3} is strongly dependent on the luminosity of the AGN.  Low luminosity AGN (log L([OIII]) $<$ 40 erg/s, left panel of Figure \ref{lo3_xs}) are only more common in pair galaxies of intermediate separation (10$<r_p<$60 kpc), by fairly modest amounts (at most a factor of $\sim$ 2) and with no significant enhancements seen in the post-mergers. Moderate luminosity AGN (40$<$ log L([OIII]) $<$ 41 erg/s, middle panel of Figure \ref{lo3_xs}) are more frequent in the closest pairs ($r_p < 10$ kpc) and post-mergers with $T_{PM}<$1 Gyr by factors of about two.  However, it is for the most powerful NLAGN (log L([OIII]) $>$ 41 erg/s, right panel of Figure \ref{lo3_xs}) that we see the greatest merger-induced enhancements, with recently coalesced galaxies exhibiting a factor of five more NLAGN than the control sample.  However, such luminous NLAGN are in the minority, so that the statistics of NLAGN as a whole (e.g. left panel of Figure \ref{agn_3}) are dominated by the weaker signal of lower luminosity events.  Figure \ref{lo3_xs} therefore demonstrates that mergers preferentially trigger luminous NLAGN.  Figure \ref{lo3_xs} also suggests that AGN of different luminosities are triggered at different times during the merger sequence.  Low luminosity AGN events occur already after the first pericentric passage whilst the galaxies continue to orbit one another in the pair phase.  Moderate luminosities are associated with the imminent merging of the pair (separations $r_p<10$ kpc) and throughout the post-merger phase.  The highest luminosities have a clear link with recent coalescence.

\begin{figure}
	\includegraphics[width=8.5cm]{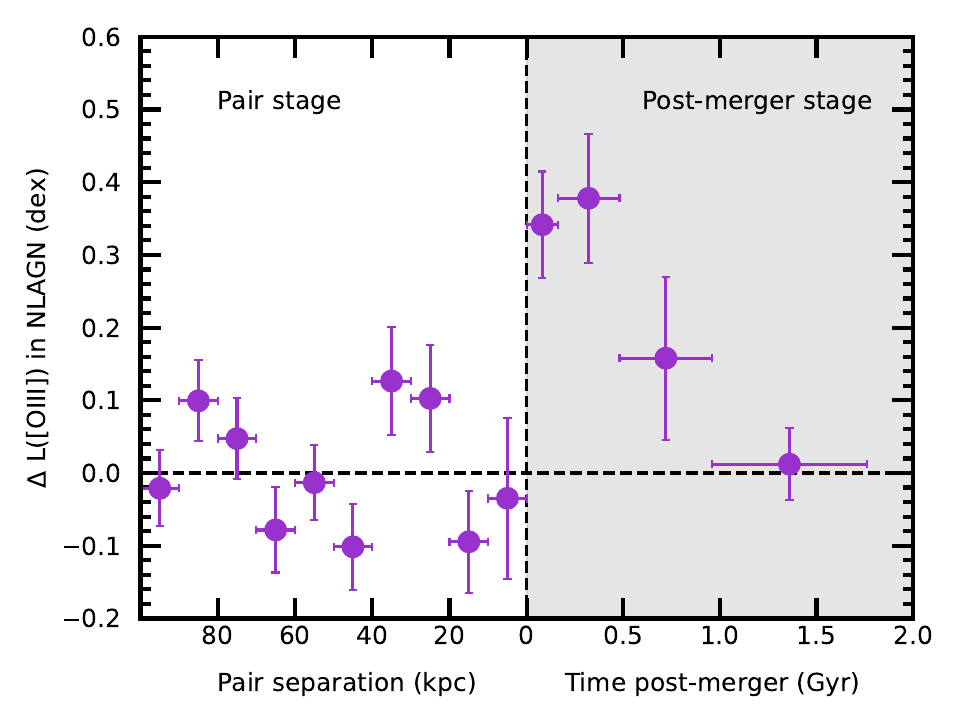}
        \caption{The enhancement in NLAGN luminosity (as measured by the luminosity of the [OIII]$\lambda$5007 line) of mergers compared with stellar mass and redshift matched control samples of non-merger NLAGN.   On average, the AGN luminosity is enhanced by a factor of 2.5 immediately after coalescence, with a statistically significant signal detected until $\sim$ 1 Gyr post-merger.}
        \label{dlo3}
\end{figure}

\smallskip

As a complementary test of our conclusion that mergers preferentially generate more luminous AGN than secularly triggered AGN, we compute the statistical difference between L([OIII]) in merger-triggered NLAGN compared with L([OIII]) in non-merger NLAGN.  In order to achieve the best possible statistical assessment of the AGN luminosity in non-mergers, we generate a new control sample, whereby the number of controls matched to each NLAGN in the merger samples is maximized and need not be the same for each galaxy\footnote{In contrast, the assessment of the fractional excess of AGN frequencies \textit{does} require the same number of controls for each merger, so the new matching scheme introduced below is not appropriate for our previous experiments.}.  Specifically, for each NLAGN host galaxy in the pair and post-merger samples, we identify all NLAGN in the control pool whose stellar mass is within $\pm$0.1 dex and redshift is within $\pm$ 0.01.  Note that in addition to the number of matches, the control pool is fundamentally different from our fiducial method because only non-merger NLAGN are considered, because we aim to compare the properties of secularly triggered AGN from those initiated by the interaction.  The control matching is deemed successful if at least five controls are found.

\smallskip

\begin{figure*}
	\includegraphics[width=9cm]{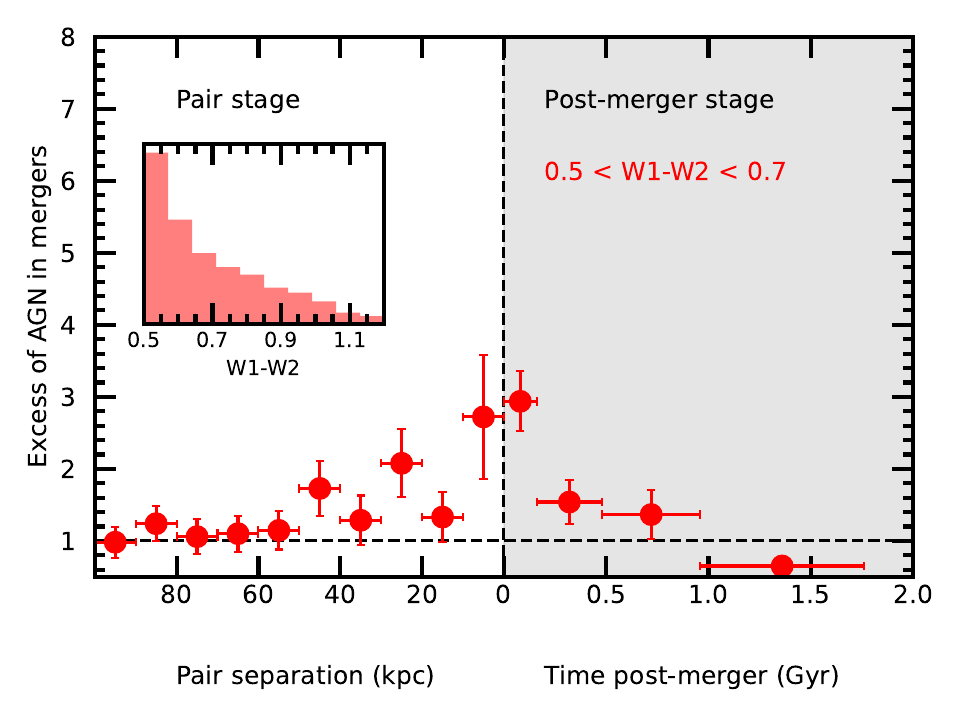}
	\includegraphics[width=9cm]{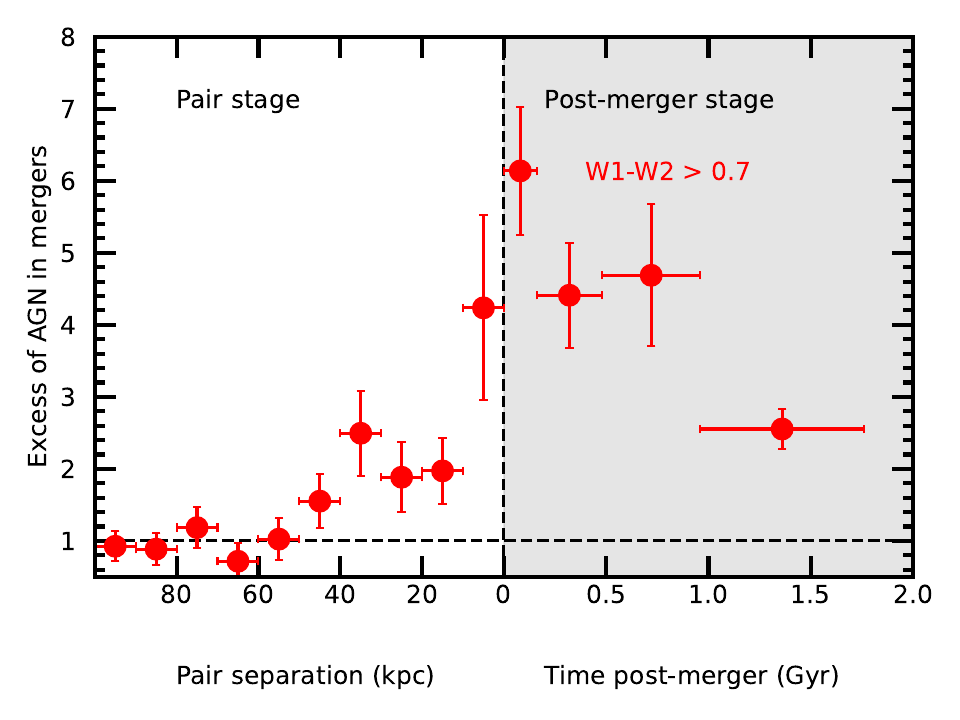}
        \caption{The excess of mid-IR AGN host galaxies as a function of $W1-W2$ colour, which correlates with AGN luminosity.   The left and right panels show AGN excess statistics for $0.5 < W1 - W2 < 0.7$ and $W1-W2 > 0.7$ respectively.  The y-axis extent is fixed for both panels in order to facilitate comparison. The inset histogram shows the $W1-W2$ distribution for the full SDSS mid-IR sample (with a lower limit of 0.5) as a reference for the distribution of the sample.}
        \label{w1w2_xs}
\end{figure*}

All NLAGN in galaxy pairs are successfully matched, typically with $>$100 controls per galaxy.  The vast majority ($>$99 percent) of post-mergers are also successfully matched, typically with $>$30 controls per galaxy (the control pool is smaller because it is limited to the UNIONS-SDSS overlap).  For the minority of post-merger NLAGN hosts with fewer than five controls the matching tolerances are increased by 0.1 dex in stellar mass and 0.01 in redshift, which yields a successful match for the remaining cases.  We then compute a $\Delta L([OIII])$ for each merging NLAGN galaxy, which is the difference (in log space) between the [OIII] luminosity of the merger and the median L([OIII]) of its controls.  Thus, the calculation of $\Delta L([OIII])$ is directly analogous to previous calculations of star formation rate enhancements ($\Delta$SFR) in mergers presented in numerous previous works (e.g. Scudder et al. 2012; Bickley et al., 2022; Ferreira et al. 2025).

\smallskip

Figure \ref{dlo3} shows the luminosity enhancement in merger-hosted NLAGN compared with non-merger NLAGN.  The signal in the pair regime is stochastic and not convincing.   However, in the post-coalescence regime, Figure \ref{dlo3} clearly shows that mergers host significantly more luminous NLAGN than are found in secular (control) NLAGNs.  The luminosity enhancement is a factor of 2.5 shortly after coalescence, declining with time, until by $\sim$1 Gyr merger-hosted NLAGN have luminosities consistent with secular NLAGN.

\smallskip

\begin{figure}
	\includegraphics[width=8.5cm]{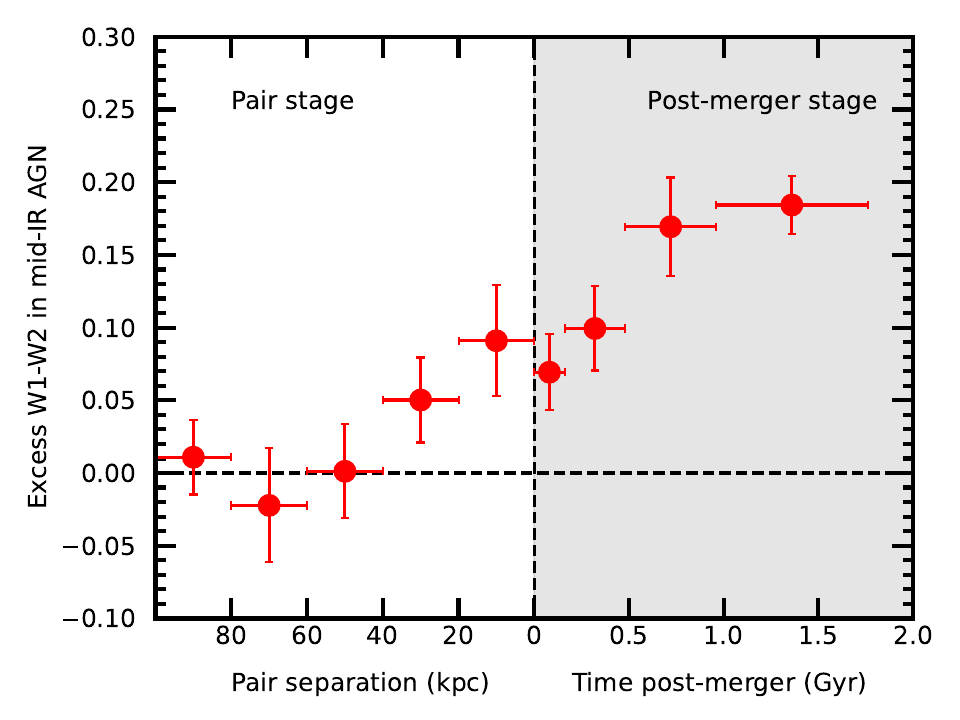}
        \caption{The enhancement of $W1-W2$ colour of mid-IR AGN in mergers compared with stellar mass and redshift matched control samples of non-merger mid-IR selected AGN.  The $W1-W2$ colour of mid-IR AGN in mergers is significantly redder than non-merger mid-IR AGN in the close pair and post-merger regimes, by up to 0.2 mag in the longest $T_{PM}$ bins.}
        \label{dw1w2}
\end{figure}

We can also investigate the role of AGN luminosity in the mid-IR selected sample. $W1-W2$ colour is an indirect measure of AGN luminosity, since it is sensitive to the relative contribution of AGN and stellar heated dust (e.g. Figure 1 of Donley et al. 2012).  Increasing the $W1-W2$ colour cut beyond our default value of 0.5 therefore selects objects with a higher fractional contribution of the AGN relative to the stellar population.  In a mass matched sample, this is equivalent to selecting higher AGN luminosities.

\smallskip

In Figure \ref{w1w2_xs} we therefore re-make the middle panel of Figure \ref{agn_3} (mid-IR AGN excess for $W1-W2>0.5$) but now imposing two different colour cuts required for a positive AGN identification\footnote{Only two bins of $W1-W2$ are used due to the very skewed distribution of colours, as can be seen in the inset histogram of the left panel of Figure \ref{w1w2_xs}.}.  In the left panel of Figure \ref{w1w2_xs} we plot the statistics for AGN whose mid-IR colour $0.5 < W1-W2 < 0.7$; the inset histogram shows the distribution of colours in the full sample for reference.  In the right panel, a much redder colour ($W1 - W2>0.7$) is required in order to be labelled as an AGN.  Figure \ref{w1w2_xs} shows that selecting more bolometrically dominant (i.e. redder) AGN leads to higher excesses.  In this case, the peak excess of mid-IR selected AGN is a factor of three for the bluer ($0.5 < W1-W2 < 0.7$) selection and a factor of six for the redder ($W1 - W2 > 0.7$) range.  Moreover, the post-coalescence time period over which the two colour cuts yield an excess differs.  Whereas the $0.5 < W1-W2 < 0.7$ shows no excess beyond $T_{PM} \sim$ 1 Gyr, the $W1 - W2 > 0.7$ shows a long-lived excess until at least 1.76 Gyr.

\smallskip

Just as we computed an excess of L([OIII]) for NLAGN ($\Delta$L([OIII]), shown in Figure \ref{dlo3}) we can compute a $W1-W2$ excess for mid-IR selected AGN.  That is, by creating a stellar mass and redshift matched sample of non-merger mid-IR AGN to the mid-IR AGN in the pairs and post-mergers, we calculate whether the latter have redder $W1-W2$ colours than the former.  The results are presented in Figure \ref{dw1w2}.  The mid-IR selected AGN in close pairs of galaxies ($r_p < $ 40 kpc) show a slightly (but statistically significant) redder colour by 0.05 -- 0.1 mag than their control sample of non-merger mid-IR AGN.  However, the colour enhancement becomes stronger in the post-coalescence regime, with mid-IR AGN in post-mergers exhibiting $W1-W2$ colours 0.1 - 0.2 mags redder than the control sample of mid-IR AGN in non-mergers throughout the post-coalescence regime.   Figure \ref{dw1w2} therefore demonstrates that mid-IR selected AGN in close pairs and post-mergers have higher fractional contributions from the AGN (i.e. the AGN is more bolometrically dominant) compared with secularly triggered AGN.

\smallskip

It is interesting to note that the greatest excess in mid-IR colour occurs in the longest $T_{PM}$ bins, even though the enhancement in AGN \textit{frequency} peaks around the time of coalescence (middle panel of Figure \ref{agn_3}).  The peak in [OIII] luminosity enhancements also occurs in the early post-merger regime (Figure \ref{dlo3}) so it might seem puzzling that the mid-IR colour excess for merger-triggered AGN is so long-lived.  We hypothesize that this might be a result of the combination of the ongoing (albeit, declining) effect of AGN triggering plus the quenching of star formation.  Using the same \textsc{mummi} merger identification method as presented here, Ellison et al. (2024) found that mergers rapidly quench star formation around the time of coalescence.  Therefore, late-stage post-mergers are less likely to be actively star forming.  In turn, this could result in galaxies with merger-induced nuclear accretion triggered at long $T_{PM}$ being more likely to be dominated by dust heated by the AGN compared with stars, compared with control galaxies of the same stellar mass.

\smallskip

Figures \ref{lo3_xs} to \ref{dw1w2} are all complementary demonstrations that mergers tend to trigger more luminous and bolometrically dominant AGN than occur through secular mechanisms.  Luminosity may also be the key to understanding the apparently puzzling differences seen in the left and middle panels of Figure \ref{agn_3}, in which the NLAGN and mid-IR selected AGN show distinct behaviours in the longest timescale bin.  In the final part of this section we address the question of why NLAGN only show an excess out to a $T_{PM} \sim 1$ Gyr, whereas mid-IR AGN (and BLAGN) are enhanced out to the largest times in our sample (1.76 Gyr post-coalescence). 

\smallskip

\begin{figure}
	\includegraphics[width=8.5cm]{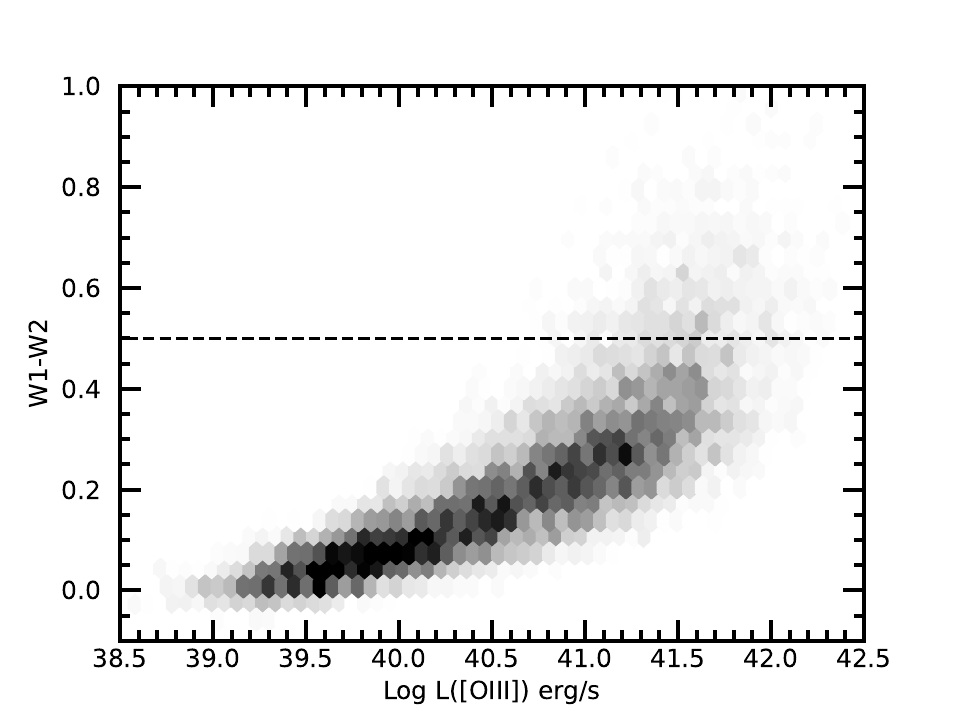}
        \caption{The distribution of $W1-W2$ colours for NLAGN in the full SDSS sample as a function of log L([OIII]). The $W1-W2>0.5$ colour cut used to identify mid-IR AGN (shown as a horizontal dashed line) tends to only select the most luminous NLAGN, and even then the selection is highly incomplete.}
        \label{lo3_w1w2}
\end{figure}

As discussed above, AGN are only identified in the mid-IR when the hot nuclear dust is bolometrically dominant over the cooler dust that is heated by stars.  As a result, not all AGN are identified in the mid-IR, as seen clearly for our sample in the Venn diagram in Figure \ref{venn_fig}.  Indeed, Goel et al. (in prep) have estimated that 80 per cent of NLAGN are missed when using a colour cut of $W1-W2>0.5$, suggesting that red mid-IR colours only identify the more luminous AGN.  In Figure \ref{lo3_w1w2} we test this idea by plotting the distribution of L([OIII]) vs. $W1-W2$ for all NLAGN in the SDSS sample; the horizontal dashed line shows the $W1-W2>0.5$ cut.  Figure \ref{lo3_w1w2} shows that only NLAGN with log L([OIII]) $\gtrapprox$ 41 erg/s have $W1-W2 > 0.5$, and even then the selection is highly incomplete with many high luminosity NLAGN falling below this mid-IR colour threshold.  The combination of the figures presented in this Section therefore allow us to understand the excesses of NLAGN and mid-IR selected AGN shown in the left and middle panels of Figure \ref{agn_3}.  Mergers preferentially trigger luminous and bolometrically dominant AGN, as seen via their [OIII] luminosities (Figures \ref{lo3_xs} and \ref{dlo3}) and by their redder WISE colours (Figures \ref{w1w2_xs} and \ref{dw1w2}).  Combined with the fact that most NLAGN have fairly low luminosities (inset in the left panel of \ref{lo3_xs}) and mid-IR selected AGN only identify a subset of the high luminosity tail of NLAGN (Figure \ref{lo3_w1w2}) the signal of triggered AGN is much more evident for mid-IR AGN, than NLAGN.  Put another way, the impact of merger triggered AGN is best seen when selecting luminous AGN, e.g. with a high L([OIII]) or with red WISE colours.  Similarly, the complete census of AGN presented in Figure \ref{agn_all_fig} is dominated by low luminosity AGN.  The results presented in this section therefore demonstrate that not only the magnitude of the AGN excess in mergers, but also the time period of which it is observed, is sensitive to the luminosity of the accretion events themselves.

\section{Discussion}

\subsection{Comparison to simulations}

We have presented the first ever observational temporal deconstruction of AGN triggering throughout the merger sequence, starting with widely separated pairs, tracing the interaction through to coalescence and 1.8 Gyr beyond.   Our observational results share several qualitative similarities with the properties of mergers in the TNG simulation, as studied by Byrne-Mamahit et al. (2024).   First, an excess of AGN is already seen in the pair phase, with increasing magnitude towards smaller projected separations, in both observations (Figure \ref{agn_3}) and simulations (Figure 3 in Byrne-Mamahit et al. 2024).  Moreover, both observations (Figure \ref{agn_3}) and simulations (Figure 6 of Byrne-Mamahit et al. 2024) find that the peak AGN excess occurs around the time of coalescence, but with a long lived influence of the merger enhancing AGN frequency for 1--2 Gyr.  Both studies are also in agreement that the AGN excess is higher for more luminous AGN (Figures \ref{lo3_xs} and \ref{dlo3} of this study and Figure 6 of Byrne-Mamahit et al. 2024), even though the AGN luminosities in the simulations are higher than the ones in our sample (assuming a bolometric correction of 142, based on Lamastra et al. 2009 and the typical range of L([OIII]) in our sample, our maximum AGN luminosities are $\sim 2 \times 10^{43}$ erg/s, compared with 10$^{44}$ for the lowest luminosity bin shown in Figure 6 of Byrne-Mamahit et al. 2024).  The TNG simulation therefore seems to broadly capture both the temporal features and luminosity dependence of merger induced nuclear accretion, despite not actually resolving the process.  In contrast, the EAGLE simulation predicts that the peak of AGN activity is significantly delayed, by about 300 Myr after coalescence (McAlpine et al. 2020), a lag that disagrees with both the NLAGN and mid-IR selected AGN statistics in this work (Figure \ref{agn_3}).

\subsection{The AGN-starburst connection}

Starbursts and AGN have long been known to go hand-in-hand, with mergers being just one possible trigger for the two processes (Sanders et al. 1988; Springel et al. 2005).  One persistent discussion in the literature has been the relative timing of the starburst and the nuclear activity, with several works suggesting a delay of a few hundred Myr between the epochs of peak formation and black hole accretion (e.g. Schawinski et al. 2009; Wild et al. 2010).  Simulations have explained this delay via dynamical and viscous lags, whereby the gas takes an extended time to reach the central supermassive black hole (Hopkins 2012; Blank \& Duschl 2016).  However, other observational works have concluded that, in mergers, starbursts and AGN occur quasi-simultaneously (Bessiere et al. 2014, 2017). 

\smallskip

Much of this previous observational work has relied on modelling stellar populations and assessing the epoch at which different physical proceseses (such as starbursts) are occuring. By quantifying the time since merger \textsc{mummi} offers a complementary tool with which to study the sequence of events that are triggered during galaxy-galaxy interactions.  In Ferreira et al. (2025) we showed that the peak of star formation occurs around the time of coalescence.  Specifically, SFRs are statistically elevated by a factor of $\sim$ 2 in the closest pairs ($r_p < 10$ kpc) and the first two post-merger bins ($T_{PM} < 0.48$ Gyr).  Using reconstructed orbits from the TNG simulation, Patton et al. (2024) calculated that pair separations decrease, on average, by 55 kpc/Gyr.   Thus we can conclude that the strongest SFR enhancements begin approximately 0.2 Gyr before coalescence and are in decline by $\sim$ 0.5 Gyr post-merger.

\smallskip

However, the enhancements in the SFR computed by Ferreira et al. (2025) are subtly different from identifying a starburst event.  That is, a galaxy can have an elevated SFR for its stellar mass, but still be forming stars at a relatively modest rate.  Moreover, Ferreira et al. (2025) explicitly exclude AGN from their sample in order that the SFR can be determined from the H$\alpha$ line.  Therefore, in order to pinpoint the chronological peak for starbursts and AGN, we take a slightly different approach.  First, we compute a SFR offset ($\Delta$SFR) from the star-forming main sequence (SFMS) for all of the galaxies (mergers and controls) in our various samples.  The $\Delta$SFR of each galaxy is computed relative to a control sample of star-forming galaxies (that define the SFMS) of the same stellar mass (within $\pm0.1$ dex) and redshift (within $\pm$0.005).  The SFRs we use are those provided in the MPA/JHU catalog, which are based on the H$\alpha$ emission line for star-forming galaxy and a calibration between the specific SFR (sSFR) and the 4000 Angstrom break (D4000) for galaxies with either a NLAGN contribution or weak H$\alpha$.  We define a starburst as a galaxy that lies at least a factor of two above the SFMS, i.e. $\Delta$SFR$>$0.3 dex.  We can then compute the excess of starbursts in the mergers and controls in exactly the same way as we have previously computed an excess of AGN.

\smallskip

\begin{figure}
	\includegraphics[width=8.5cm]{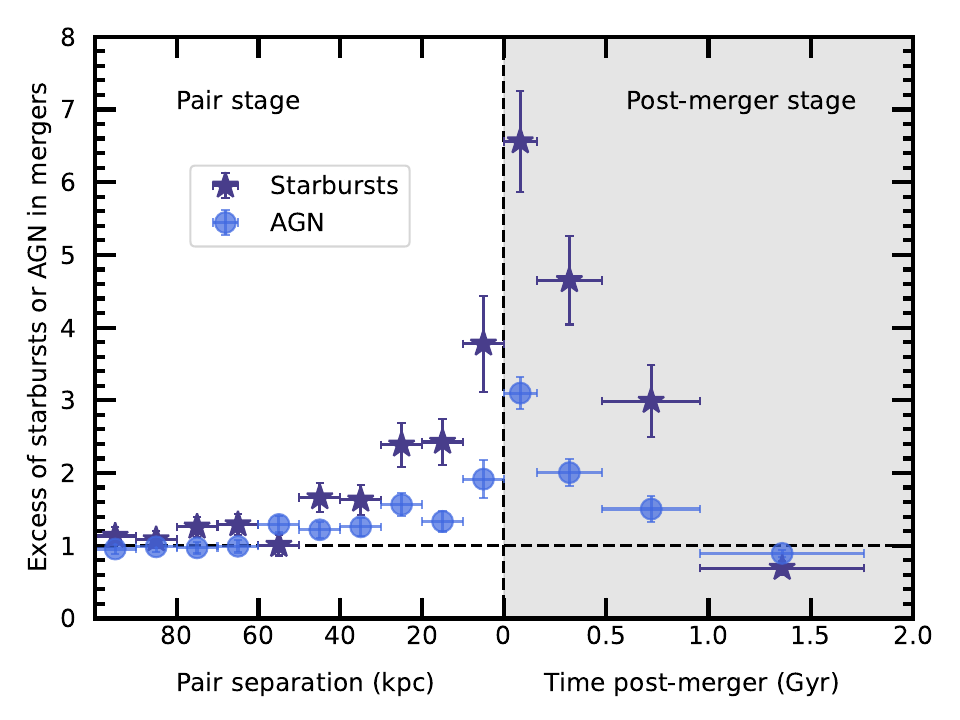}
        \caption{The excess of both starbursts (star symbols) and AGN (circles) identified by any of the diagnostics used in this paper (NLAGN, BLAGN and mid-IR colours, reproduced from Figure \ref{agn_all_fig}) in the pairs (left side of the diagram) and post-mergers (right side of the diagram) compared with the control sample. The horizontal dashed line shows an excess of one, i.e. the same fraction in mergers and controls.   The pair sample is binned in units of projected separation (kpc) whereas the post-mergers are binned by the time post-merger, with x-axis error bars indicating the width of the $r_p$ or $T_{PM}$ bin.  The peak in the starburst excess occurs contemporaneously with the peak in AGN triggering, i.e. immediately after coalescence at $T_{PM}<0.16$ Gyr.}
        \label{SB_AGN}
\end{figure}

In Figure \ref{SB_AGN} we show the excess of starbursts as a function of time through the merger sequence.  The greatest excess in the occurence of starbursts is immediately after coalescence, at $T_{PM}<$0.16 Gyr.  For comparison, we also re-produce in Figure \ref{SB_AGN} the AGN excess considering all of the diagnostics (NLAGN, BLAGN, mid-IR AGN) shown previously in Figure \ref{agn_all_fig}.  Figure \ref{SB_AGN} shows that the peak in the triggering of starbursts and AGN occurs contemporaneously, immediately after coalescence.  However, given the relative coarseness of the time bins, our results nonetheless permit a lag between these two physical processes, but by no more than $\sim$ 150 Myr.  We also stress that our results speak only to the statistical peak of starbursts and AGN across the population of mergers in our sample, and do not constrain the time lag between a starburst and nuclear accretion in a given galaxy.  Nonetheless, our results clearly demonstrate that the peak of activity for both star formation and nuclear accretion occur around the time of coalescence on a population-wide basis.

\smallskip

\subsection{The luminosity dependence of merger-triggered AGN}

We have shown in Figure \ref{lo3_xs} that the excess frequency of AGN depends on the luminosity regime in which the statistics are assessed.  For low luminosity NLAGN the signal is weak, and the enhancement only seen in a fairly narrow range of pair separations (10$<r_p<$40 kpc).  In the moderate luminosity regime, a fairly constant factor of two excess of NLAGN is seen in the closest pairs and throughout the post-merger epoch until $T_{PM} \sim$ 1 Gyr. Conversely, for the most luminous AGN there is a clear link to the immediate post-coalescence regime.  We further show in Figure \ref{dlo3} that NLAGN in post-mergers are more luminous than NLAGN in a mass and redshift matched control sample of non-mergers.  Taken together, our results support the idea that mergers not only trigger AGN, but that these events are more luminous than secularly triggered accretion events.  Our results are consistent with studies that use statistical samples drawn from large cosmological simulations (McAlpine et al. 2020; Byrne-Mamahit et al. 2023, 2024).  Many other observational studies have also concluded that low redshift mergers preferentially have higher luminosities and/or host more bolometrically dominant AGN, as measured either with [OIII] (e.g. Alonso et al. 2007; Ellison et al. 2019; Pierce et al. 2022, 2023; Bickley et al. 2023), X-rays (e.g. Hou et al. 2020; La Marca et al. 2024), mid-IR (Satyapal et al. 2014) or radio (Ramos Almeida et al. 2011). Conversely, the recent study of Villforth (2023), which made the most complete compilation of multi-wavelength data currently available in the literature, found no link between AGN fraction and sample luminosity.

\smallskip

We suggest that the reconciliation of previous studies on the luminosity dependence of AGN triggering can be rooted in a few compounding causes.  Several issues are related to sample selection.  Whilst the majority of recent studies of large samples at \textit{low redshift} have found a merger-AGN connection, the same is not true at higher redshift (Kocevski et al. 2012; Shah et al. 2020; Silva et al. 2021; Lambrides et al. 2021; Dougherty et al. 2024).  Therefore, samples that investigate the luminosity dependence over large redshift ranges may be more challenged in extracting a signal (if one is there).  Furthermore, multi-wavelength studies, including ours (see also recent studies by Gao et al. 2020; Bickley et al. 2023, 2024a; La Marca et al. 2024) have clearly demonstrated that results depend on selection method.  This is partly because mid-IR selection tends to identify powerful, obscured AGN (e.g. Satyapal et al. 2017; Blecha et al. 2018) but also because NLAGN tend to be less luminous than those selected by other methods (e.g. Figure \ref{lo3_w1w2} and Bickley et al. 2024a).  Finally, the application of bolometric correction factors required to homogenize diverse samples introduces further uncertainty and may potentially erase a real signal.  Our work has underscored a further important factor to the issues associated with the study of the role of luminosity -- timescale of the interaction.  Figure \ref{dlo3} shows that large enhancements in the AGN luminosity are generally associated with the post-merger phase, in agreement with previous studies that have had a more limited view of the post-merger regime (Satyapal et al. 2014; Ellison et al. 2019; Bickley et al. 2023).  We conclude that studies with large statistical samples, selected in a consistent way (e.g. wavelength, timescale), at low redshift, with well-matched control samples usually find a link between AGN power and merger status.

\section{Summary}\label{summary_sec}

We have identified AGN for a sample of close galaxy pairs and post-mergers, whose time since coalescence is predicted by \textsc{mummi}, in order to make the first ever assessment of enhanced nuclear accretion throughout the merger sequence.  AGN are identified via three complementary techniques: narrow and broad emission lines, and mid-IR colours.  A matched control sample of non-interacting galaxies allows us to compare the fraction of AGN in mergers with a reference sample and compute an AGN excess. Our main conclusions are as follows.

\begin{itemize}

\item  \textbf{The AGN excess peaks immediately after coalescence, at $T_{PM} < 0.16$ Gyr.} By combining all three AGN diagnostics together, in order to have the most complete census available to us, we find that AGN enhancements begin to be detectable already during the pair phase, for projected separations $r_P <$ 60 kpc.  However, the peak in AGN excess is seen immediately after coalescence, see Figure \ref{agn_all_fig}.  The peak AGN excess at $T_{PM} < 0.16$ Gyr is also seen for all of the AGN diagnostics when considered separately, with the greatest enhancement seen for mid-IR selected AGN (Figure \ref{agn_3}).

\item \textbf{Nuclear obscuration evolves through the merger sequence.}  BLAGN are only visible when the line of sight to the nuclear region is relatively unobscured by dust.  The lack of enhancement (and even a mild deficit) of BLAGN seen in the pair phase is therefore consistent with enhanced obscuration and/or higher dust covering fractions as a result of tidally induced gas inflows.  In the post-merger phase, radiation pressure from the AGN re-distributes this dust and the BLAGN becomes visible (right hand panel of Figure \ref{agn_3}), resulting in an increased fraction of unobscured AGN in the late coalescence regime (Figure \ref{unobs_agn}).

\item \textbf{The enhancement in the AGN fraction is long-lived.}  Different AGN diagnostics are sensitive to accretion events of varied characteristics, for example in terms of luminosity, dust and accretion rate.  This is reflected in different quantitative levels of statistical enhancement.  However, for both BLAGN and mid-IR selected, a statistical excess is seen out to the longest timescales represented in our sample (1.76 Gyr post-coalescence), see Figure \ref{agn_3}.  Therefore, compared with the enhancements in SFR, which appear to return to normal levels by $T_{PM} \sim 1$ Gyr (Ferreira et al. 2025), nuclear accretion experiences a more protracted boost from the merger.  
  
\item \textbf{The greatest AGN enhancements are seen for the most luminous events.}  By using either L([OIII]) (Figure \ref{lo3_xs}) as a measure of AGN luminosity or $W1-W2$ colour (Figure \ref{w1w2_xs}) as an indicator of fractional AGN contribution, we find that the AGN excess is greatest for the most powerful/dominant AGN.  Our results support the hypothesis that the most luminous AGN are more likely to be mergers.

\item \textbf{Merger induced NLAGN are more luminous than secular NLAGN.} By comparing the [OIII] luminosities of NLAGN in mergers to NLAGN in non-mergers we find the former to be more luminous by a factor of $\sim$ 2.5 (Figure \ref{dlo3}).  

\item \textbf{The peak in the excess of starbursts occurs contemporaneously with the peak in AGN enhancement. }  Starbursts (defined here as galaxies with SFRs at least a factor of two above the SFMS) also peak at $T_{PM}<0.16$ Gyr (Figure \ref{SB_AGN}).  Any lag between the triggering of these two processes must therefore be less than $\sim$ 150 Myr.
 
\end{itemize}

Many of our results are in excellent qualitative agreement with conclusions drawn from cosmological simulations.  For example, Byrne-Mamahit et al. (2024) have used the Illustris-TNG simulation to show that the AGN enhancement peaks at coalescence and shows a statistical excess until $T_{PM} \sim$ 1.5 Gyr.  As we have also seen in the observations, the excess in AGN in the simulations is also luminosity dependent.    Moreover, again using Illustris-TNG, Hani et al. (2020) found a much shorter timescale for SFR enhancement of $\sim$ 500 Myr.  Although this is somewhat shorter than the timescale for SFR enhancements that we find in the observations ($\sim$ 1 Gyr; Figure \ref{SB_AGN} of this paper and Ferreira et al. 2025) both simulations and observations alike agree qualitatively that post-merger AGN activity is more persistent than enhancements in SFR.  However, neither observations nor simulations are able to assign \textit{all luminous} AGN to mergers (Byrne-Mamahit et al. 2023, 2024).  Higher time and spatial resolution in the simulations may be crucial to tackling this outstanding question.

\end{document}